\documentclass[twocolumn%
,secnumarabic%
,amssymb,aps,pra,nobibnotes,superscriptaddress]{revtex4-1}
\usepackage{amsmath}
\usepackage{graphicx}
\usepackage{ifpdf}
\usepackage{color}
\usepackage{xcolor}
\expandafter\ifx\csname package@font\endcsname\relax\else
 \expandafter\expandafter
 \expandafter\usepackage
 \expandafter\expandafter
 \expandafter{\csname package@font\endcsname}%
\fi
\DeclareRobustCommand\substyle{\name@idx{document substyle}}%
\DeclareRobustCommand\classoption{\name@idx{document class option}}%
\DeclareRobustCommand\classname{\name@idx{document class}}%
\def\name@idx#1#2{%
 {\ttfamily#2}%
 \index{#2\space#1=\string\ttt{#2}\space#1}\index{#1>#2=\string\ttt{#2}}%
}%

\begin{document}
\title{Transmission Line Model for Materials with Spin-Momentum Locking}%
\author{Shehrin Sayed}\email{ssayed@purdue.edu}
\affiliation{School of Electrical and Computer Engineering, Purdue University, West Lafayette, IN 47907, USA}
\author{Seokmin Hong}
\affiliation{Center for Spintronics, Korea Institute of Science and Technology, Seoul 02792, Republic of Korea.}\email{shong@kist.re.kr}
\author{Supriyo Datta}\email{datta@purdue.edu}
\affiliation{School of Electrical and Computer Engineering, Purdue
	University, West Lafayette, IN 47907, USA}

\begin{abstract}
We provide a transmission line representation for channels exhibiting spin-momentum locking (SML) which can be used for both time-dependent and steady-state transport analysis on a wide variety of materials with spin-orbit coupling such as topological insulators, heavy metals, oxide interfaces, and narrow bandgap semiconductors. This model is based on a time-dependent four-component diffusion equation obtained from the Boltzmann transport equation assuming linear response and elastic scattering in the channel. We classify all electronic states in the channel into four groups ($U^+$, $D^+$, $U^-$, and $D^-$) depending on the spin index (up ($U$), down ($D$)) and the sign of the $x$-component of the group velocity ($+,-$) and assign an average electrochemical potential to each of the four groups to obtain the four-component diffusion equation. For normal metal channels, the model decouples into the well-known transmission line model for charge and a time-dependent version of Valet-Fert equation for spin. We first show that in the steady-state limit our model leads to simple expressions for charge-spin interconversion in SML channels in good agreement with existing experimental data on diverse materials. We then use the full time-dependent model to study spin-charge separation in the presence of SML ($p_0\neq 0$), a subject that has been controversial in the past. Our model shows that the charge and spin signals travel with two distinct velocities resulting in well-known spin-charge separation which is expected to persist even in the presence of SML. However, our model predicts that the lower velocity signal is purely spin while the higher velocity signal is largely charge with an additional spin component proportional to $p_0$, which has not been noted before. Finally, we note that our model can be used within standard circuit simulators like SPICE to obtain numerical results for complex geometries.
\end{abstract}
\maketitle

\section{Introduction}
\textit{Background}: Transport properties in materials with spin-orbit coupling (SOC) are of great interest for potential spintronic applications, especially because of the unique spin-momentum locking (SML) observed in diverse classes of materials such as topological insulator (TI) \cite{Hasan_RMP_2010, Zhang_RMP_2011, Wang_SPIN_2016}, heavy metals \cite{Miron_Nat_2011, Ohno_APL_2011, Ralph_APL_2012, Ralph_Science_2012, Felser_NatComm_2015}, oxide interfaces \cite{FertLAOSTO2016, Songe1602312}, and narrow bandgap semiconductors \cite{Johnson_PRB_2000, Koo_JAppPhys_2012, SilsbeeJPhys2004}. There has been an immense effort to model the interplay between spin and charge in such materials using time-dependent classical \cite{CosimoPRB2017} or quantum Boltzmann equation \cite{SinovaPRB2012,SinovaPRL2013}, nonequilibrium Green's function \cite{HalperinPRL2004, BurkovPRB2004, BurkovPRL2010, ZainuddinPRB2011, Hong_PRB_2012}, phenomenological equations coupled to magnet dynamics \cite{TserkovnyakPRB2014}, and time-independent diffusion equation used to explain bulk spin Hall effect \cite{BauerPRB2013}.

\textit{Four-Component Diffusion Equation}: In this paper, we propose a time-dependent four-component diffusion equation that can be used for transport analysis on multi-contact based structures implemented with materials exhibiting SML. The model is obtained from the Boltzmann transport equation assuming linear response and elastic scattering processes in the channel. The basic approach is to assign one electrochemical potential $\mu(\vec{p},s)$ to each of the eigenstates ($\vec{p},s$) where $\vec{p}$ is the momentum confined in the $z$-$x$ plane and $s=\pm1$ is the spin index with $+1$ and $-1$ denoting the up ($U$) and the down ($D$) spins with respect to the spin quantization axis  $\hat{y}\times\left(\vec{p}-q\vec{A}\right)$ ($\vec{A}$ is the vector magnetic potential).

We then classify the eigenstates into four groups ($U^+$, $D^+$, $U^-$, and $D^-$) based on the spin index ($U$, $D$) and the sign of the $x$-component of the group velocity ($+,-$) and define an average electrochemical potential corresponding to the each of the four groups resulting in a four-component diffusion equation. This can be viewed as an extension of the Valet-Fert equation which uses two electrochemical potentials for $U$ and $D$ states \cite{Valet_Fert_1993}. The four-component diffusion equation in steady-state reduces to our prior model \cite{Sayed_SciRep_2016, Hong_SciRep_2016} that we used to predict a unique three resistance state on SML materials with two FM contacts in a multi-terminal spin valve structure \cite{Sayed_SciRep_2016}. The prediction has been observed recently on Pt \cite{Pham_NanoLett_2016, Pham_APL_2016} and InAs \cite{Koo_New} up to room temperature. We expect the prediction to be observed on any channel exhibiting SML.

In our generalized view, $U^+$ (and $U^-$) states have same number of modes $M$ (and $N$) as $D^-$ (and $D^+$) states due to the time reversal symmetry. The degree of SML in our model is given by \cite{Sayed_SciRep_2016, Hong_SciRep_2016}
\begin{equation}
\label{degSML}
p_0=\dfrac{M-N}{M+N},
\end{equation}
where $M$ and $N$ are evaluated at Fermi energy for zero temperature and in general require thermal averaging. For normal metal (NM) channels $p_0=0$ i.e. $M=N$. For a perfect topological insulator (TI) $N=0$ leading to $p_0=1$, however, $p_0$ gets effectively lowered by the presence of parallel channels. $p_0$ has been quantified for different TIs by a number of groups \cite{JonkerNatNano2014, KLWangNanoLett2014, DasNanoLett2015, SamarthPRB2015, YPChenSciRep2015, Samarth_PRB_2015, YoichiPRB2016} by measuring the charge current induced spin voltage using a ferromagnetic (FM) contact, motivated by a theoretical proposal \cite{Hong_PRB_2012}. For a Rashba channel with a coupling coefficient $\alpha_R$, $p_0 \approx \alpha_R k_F /(2E_F)\ll 1$ \cite{SilsbeeJPhys2004, Hong_PRB_2012} which can be quantified with similar spin voltage measurements \cite{Johnson_PRB_2000,Koo_JAppPhys_2012}. $k_F$ and $E_F$ are the Fermi wave vector and Fermi energy respectively. Recently, spin voltage measurements have been reported on heavy metals like platinum \cite{Pham_NanoLett_2016, Pham_APL_2016} and gold \cite{AppelbaumPRB2016}. These experiments can also be quantified by $p_0$, though the underlying mechanism is subject to active debate \cite{StilesPRB2013, SinovaRevModPhys2015, HoffmannIEETMAG2013} and could involve a bulk spin Hall effect \cite{Ralph_Science_2012, BauerPRB2013, WangPRL2014} or interface Rashba-like channel \cite{Felser_NatComm_2015, Saitoh_SciRep_2015, HoeschPRB2004, Tamai_PRB_2013}.

\textit{Transmission Line Model}: We translate our four-component ($U^+$, $D^+$, $U^-$, and $D^-$) semiclassical model into a transmission line model with two component (charge and $z$-component of spin) voltages and currents where the coupling between charge and spin in a SML channel is characterized by $p_0$ in Eq. \eqref{degSML}. The model is compatible within a standard circuit simulator tool like Simulation Program with Integrated Circuit Emphasis (SPICE) which will enable straightforward analysis of complex geometries. The transmission line model is a new addition to our multi-physics spin-circuit framework \cite{ModApp} which has been previously used to explain experiments and evaluate spin-based device proposals \cite{Camsari_SciRep_2015, Sayed_SciRep2_2016}. 

For NM channels (i.e. $p_0=0$), the proposed transmission line model decouples into the well-known model for charge that has been previously used to analyze transport in quantum wires \cite{Burke_TNANO_2002, Burke_TNANO_2003, Sayeef_IEEE_2005} and a time-dependent version of Valet-Fert equation \cite{Valet_Fert_1993} for spin. For SML channels (i.e. $p_0\neq 0$), our model lead to several prior results on charge-spin interconversion in the steady-state limit \cite{Hong_PRB_2012, Hong_SciRep_2016} that have been previously used by a number of experimental groups \cite{JonkerNatNano2014, KLWangNanoLett2014, DasNanoLett2015, SamarthPRB2015, YPChenSciRep2015, Samarth_PRB_2015, YoichiPRB2016, Koo_New} to quantify their spin voltage measurements using potentiometric ferromagnetic contacts. We further derive a simple expression and present SPICE simulation results for a parameter that has been widely used to quantify the inverse Rashba-Edelstein effect (IREE) in 2D channels, which are in good agreement with existing experiments \cite{Fert_NatComm_2016,IssaCuBi2016,FertLAOSTO2016,SmarthPRL2016} on diverse materials.

We then use the full time-dependent transmission line model to study the spin-charge separation in the presence of SML in materials with SOC, a subject that has been controversial in the past (see, for example,  \cite{BalseiroPRL2002, CalzonaPRB2015, Stauber_PRB_2013,BarnesPRL2000}). Our model suggests that depending on the channel cross-section, the charge signal can travel faster than the spin signal resulting in spin-charge separation which is well-known for materials without SOC \cite{HalperinJAP2007, PoliniPRL2007, SchroerPRL2014, Burke_TNANO_2002} based on the Luttinger liquid theory. We argue using our model that the spin-charge separation persists even in the SOC materials exhibiting SML (i.e. $p_0\neq0$). Similar arguments have been made in the past considering the presence of spin-orbit coupling (SOC) \cite{BalseiroPRL2002, CalzonaPRB2015, Stauber_PRB_2013} although there exists counter arguments that the presence of SOC destroys the spin-charge separation \cite{BarnesPRL2000}. However, we predict that the high velocity charge signal in SML channels accompanies an additional spin component proportional to $p_0$ having the same velocity as the charge, which has not been discussed before.

Note that the proposed model does not take into account the effects such as spin precession involving the off-diagonal elements of the density matrix which we assume to be negligible. An extension of this model to include $x$ and $y$ components of spin could possibly address such issues, as done earlier for materials without SOC (see \cite{Camsari_SciRep_2015}, and references therein). The assumptions made to derive the model have been discussed in detail in Section \ref{sec_semi}. Several predictions from our model for steady-state \cite{Sayed_SciRep_2016} have already received support from experiments \cite{Pham_APL_2016,Pham_NanoLett_2016,Koo_New} suggesting that the assumptions are within the reasonable limits. The assumptions can be revisited as the field evolves leading to revised model parameters, but the basic model should remain valid.

\textit{Outline}: The paper is organized as follows. In Section II, we describe the transmission line model for SML channels and show that special cases lead to prior well-known models. In Section III, we derive several results on charge-spin interconversion from our transmission line model in steady-state and present comparison with SPICE simulations using the full model. We obtain a simple expression for a parameter that has been widely used to quantify IREE and show that it is in good agreement with available experiments on diverse materials. In Section IV, we study the spin-charge separation in terms of spin and charge signal velocities obtained from our time-dependent transmission line equations. We show that the separation persists even in SML channels, however, in SML channels there exists an additional spin component accompanied by the high velocity charge signal.  In Section V, we derive the transmission line model starting from the Boltzmann transport equation with all the assumptions clearly stated. We discuss different scattering mechanisms in the channels and their effects on charge and spin transport. Finally, in Section VI, we end with a brief summary.

\section{Transmission Line Model}

\subsection{Model Description}
\begin{figure}
	\includegraphics[width=0.47 \textwidth]{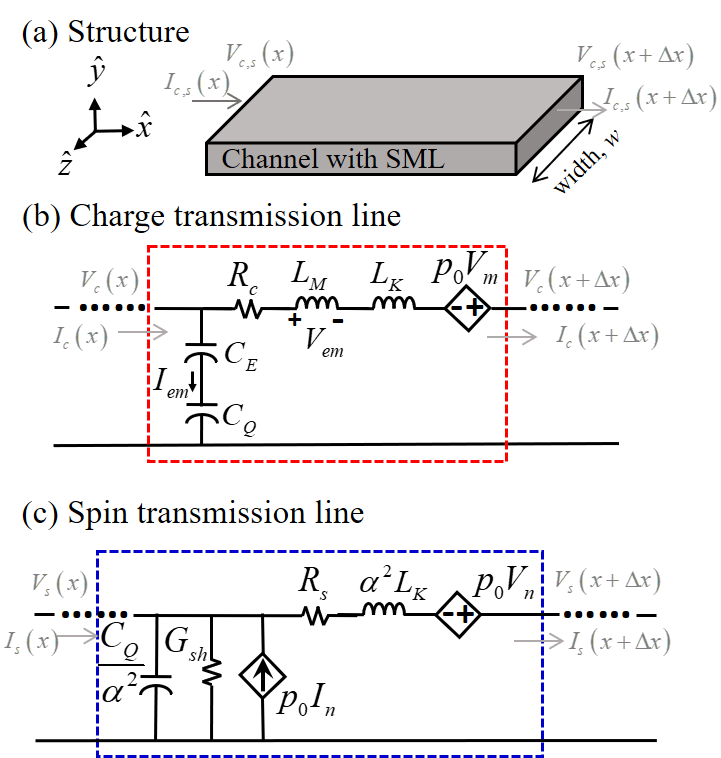}
	\caption{(a) Structure and corresponding axes of the channel with spin-momentum locking (SML) under consideration, for which a transmission line model is derived. The model has two components: (b) charge (corrseponds to Eq. \eqref{charge_TL}) and (c) spin (corrseponds to Eq. \eqref{spin_TL}), with coupling between them described by degree of SML $p_0$ (see Eq. \eqref{degSML}). Coupling between charge and spin are modeled by dependent voltage and current sources with $V_m=\eta_c V_s$, $V_n=\eta_s V_c+r_m I_{em}$, and $I_n=\gamma_s I_c - g_m V_{em}$.}\label{1}
\end{figure}

We consider the structure and axes in Fig. \ref{1}(a) to derive the transmission line model. The model has two components: charge and $z$-component of spin with coupling between them characterized by $p_0$ in Eq. \eqref{degSML}. The charge model is given by
\begin{equation}
\label{charge_TL}
\begin{aligned}
&\left(\dfrac{1}{C_E}+\dfrac{1}{C_Q}\right)^{-1}\;\dfrac{\partial}{{\partial t}}{V_c} =  - \dfrac{\partial}{{\partial x}}I_c,\\
&\left(L_K+L_M\right)\;\dfrac{\partial}{{\partial t}}I_c + R_c\,I_c =  - \dfrac{\partial}{\partial x}{V_c} + {p_0}{\eta_c}{V_s},
\end{aligned}
\end{equation}
where  $I_c$ and $V_s$ are charge current and voltage along $\hat{x}$-direction, $C_E$ and $C_Q$ are the electrostatic and quantum capacitances per unit length, $L_M$ and $L_K$ are the magnetic and kinetic inductances per unit length, and $R_c$ is the charge resistance per unit length.

The spin model is given by
\begin{equation}
\label{spin_TL}
\begin{aligned}
&\dfrac{C_Q}{\alpha^2}\dfrac{\partial}{{\partial t}}{V_s} + G_{sh}{V_s} + p_0 g_m V_{em} =  - \dfrac{\partial}{{\partial x}}{I_s} + p_0 \gamma_s I_c,\\
&\alpha^2{L_K}\dfrac{\partial}{{\partial t}}{I_s} + {R_s}{I_s} - p_0 r_m I_{em} =  - \dfrac{\partial}{{\partial x}}{V_s} + p_0 \eta_s {V_c},
\end{aligned}
\end{equation}
where $I_s$ and $V_s$ are spin current and voltage along $\hat{x}$-direction with spin polarization along the $\hat{z}$-direction. In this discussion, $\hat{y}$-direction is out-of-plane. Here, $\alpha=2/\pi$ is an angular averaging factor, $R_s$ is the spin resistance per unit length of the channel and $G_{sh}$ is the shunt conductance per unit length that captures the spin lost in the channel due to the spin relaxation. Detailed derivation of Eqs. \eqref{charge_TL} and \eqref{spin_TL} from the Boltzmann transport equation will be discussed in Section \ref{sec_semi} with clearly stated assumptions.

Distributed circuit models for charge and spin are shown in Fig. \ref{1}(b) and (c), which are based on Eqs. \eqref{charge_TL} and \eqref{spin_TL} respectively. The dependent sources proportional to $p_0$ represent charge-spin inter-coupling between the two models. The dependent source parameters in Fig. \ref{1} are given by
\begin{subequations}
	\begin{equation}
	\label{dep_Vm}
	V_m=\eta_c V_s,
	\end{equation}
	\begin{equation}
	\label{dep_Vn}
	V_n=\eta_s V_c+r_m I_{em},
	\end{equation}
	\begin{equation}
	\text{and,}\;\, I_n=\gamma_s I_c - g_m V_{em}.
	\end{equation}
\end{subequations}


The parameters of Eqs. \eqref{charge_TL} and \eqref{spin_TL} are given by
\begin{subequations}
	\label{params}
	\begin{alignat}{10}
	\label{cq}
	&C_Q = \dfrac{2}{R_B \left| {\left\langle {{v_x}} \right\rangle } \right|},\\
	\label{lk}
	&L_K=\dfrac{R_B}{2 \left| {\left\langle {{v_x}} \right\rangle } \right|},\\
	\label{rc}
	&R_c = \dfrac{R_B}{\lambda},\\
	\label{rs}
	&R_s=\dfrac{\alpha^2 R_B}{\lambda_0},\\
	\label{gsh}
	&G_{sh}=\dfrac{4}{\alpha^2 R_B\lambda_s},\\
	&g_m=\dfrac{2}{\alpha R_B},\\
	&r_m=\dfrac{\alpha}{\left| {\left\langle {{v_x}} \right\rangle } \right|C_E},\\
	\label{ets}
	&\eta_s=\dfrac{{2\alpha}}{{{\lambda _0}}},\\
	\label{etc}
	&\eta_c=\dfrac{2}{\alpha {\lambda_r}},\\
	\label{gamas}
	&\gamma_s=\dfrac{{2}}{{{\alpha\lambda_t}}},\\
	\label{bal_res}
	\text{and,}\;\;&R_B=\dfrac{h}{q^2}\dfrac{1}{M+N}.
	\end{alignat}
\end{subequations}
Here, $\lambda$, $\lambda_0$, and $\lambda_s$ are three distinct mean free paths that determine $R_c$, $R_s$, and $G_{sh}$ respectively. $|\langle v_x\rangle|$ is the the magnitude of the thermally averaged electron velocity $\langle{v}_x\rangle$. $\eta_c$ represents spin to charge conversion coefficient and $\eta_s,\gamma_s$ represent charge to spin conversion coefficients. These coefficients depend on different scattering mechanisms in the channel, which will be discussed later in Section \ref{sec_semi}. $r_m$ and $g_m$ are transient charge-spin coupling coefficients which are in the units of resistance per unit length and conductance per unit length respectively. 


$R_B$ is the ballistic resistance of the channel, $h$ is the Planck's constant, and $q$ is electron charge. $R_B$ is inversely proportional to the total number of modes ($M + N$) in the channel which represents a material property and does not imply ballistic transport. The models and related results discussed in this paper are valid all the way from ballistic to diffusive regime of operation.

\begin{figure}
	\includegraphics[width=0.49 \textwidth]{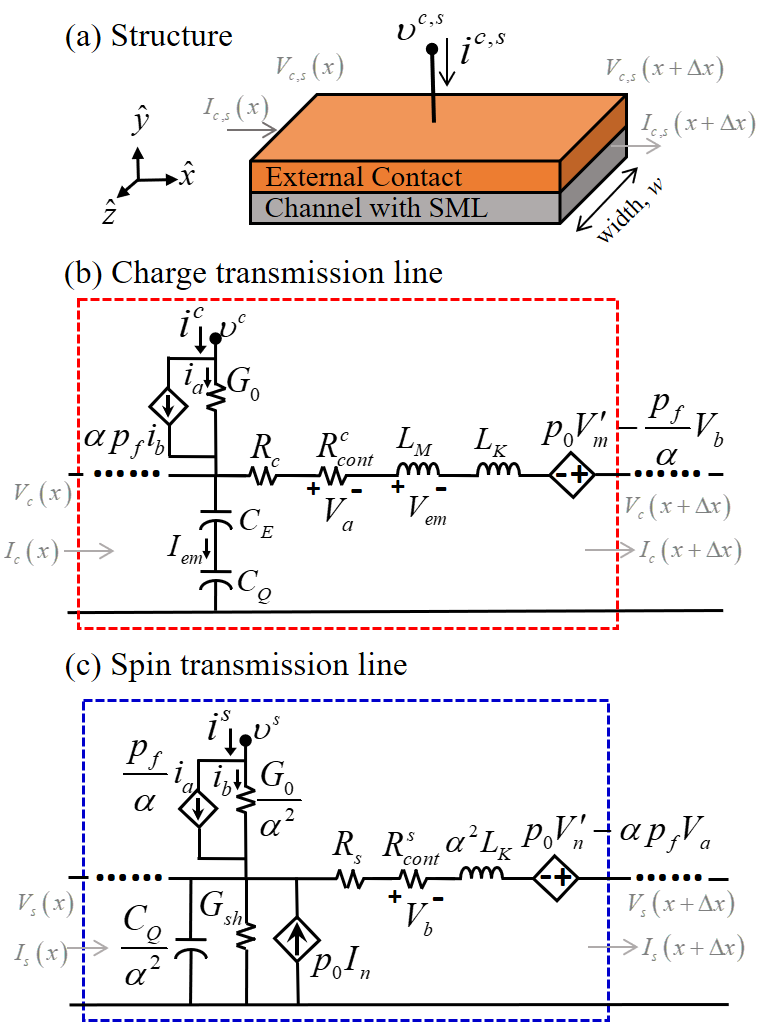}
	\caption{(a) Structure of the spin-momentum locked (SML) channel in the presence of an external contact. Both (b) charge and (c) spin transmission line models are modified as compared to Fig. \ref{1}, which now correspond to Eqs. \eqref{charge_TL_cont} and \eqref{spin_TL_cont} respectively. The dependent sources are $V_m '=\eta_c V_s+\frac{G_0 R_B}{2}\left(\frac{v_s}{\alpha}+p_f v_c\right)$, $V_n '=\eta_s V_c+r_m I_{em}+\frac{G_0 R_B}{2}\left(\alpha v_c+p_f v_s\right)$, and $I_n=\gamma_s I_c - g_m V_{em}$. Note that the model reduces to that shown in Fig. \ref{1} in the limit $G_0 \rightarrow 0$.}\label{3}
\end{figure}

\subsection{Presence of an External Contact}
In the presence of an external contact on the channel (see Fig. \ref{3}(a)), the charge model in Eq. \eqref{charge_TL} is modified as
\begin{equation}
\label{charge_TL_cont}
\begin{aligned}
&\left(\dfrac{1}{C_E}+\dfrac{1}{C_Q}\right)^{-1}\;\dfrac{\partial}{{\partial t}}{V_c} =  - \dfrac{\partial}{{\partial x}}I_c+i^c,\\
&\left(L_K+L_M\right)\;\dfrac{\partial}{{\partial t}}I_c + R_c\,I_c =  - \dfrac{\partial}{\partial x}{V_c} + {p_0}{\eta_c}{V_s}+\Delta v^c,
\end{aligned}
\end{equation}
and the spin model in Eq. \eqref{spin_TL} is modified as
\begin{equation}
\label{spin_TL_cont}
\begin{aligned}
&\dfrac{C_Q}{\alpha^2}\dfrac{\partial}{{\partial t}}{V_s} + G_{sh}{V_s} + p_0 g_m V_{em} =  - \dfrac{\partial}{{\partial x}}{I_s} + p_0 \gamma_s I_c+i^s,\\
&\alpha^2 {L_K}\dfrac{\partial}{{\partial t}}{I_s} + {R_s}{I_s} - p_0 r_m I_{em} =  - \dfrac{\partial}{{\partial x}}{V_s} + p_0 \eta_s {V_c}+\Delta v^s,
\end{aligned}
\end{equation}
where $i^c$ and $i^s$ represent charge and spin currents entering into the channel per unit length from the external contact, $\Delta v^c$ and $\Delta v^s$ represent the change in channel charge and spin voltages per unit length in the region under the external contact. They are given as
\begin{equation}
\label{contact_charge_spin}
\left\{ {\begin{array}{*{20}{c}}
	{{i^c}}\\
	{{i^s}}
	\end{array}} \right\} = {G_0}\left[ {\begin{array}{*{20}{c}}
	1&{\dfrac{p_f}{\alpha}}\\\\
	\dfrac{p_f}{\alpha} &{\dfrac{1}{\alpha^2}}
	\end{array}} \right]\left\{ {\begin{array}{*{20}{c}}
	{{v_c} - {V_c}}\\
	{{v_s} - {V_s}}
	\end{array}} \right\}, \text{ and}
\end{equation}
\begin{equation}
\label{contact_charge_spin1}
\begin{array}{l}
\left\{ {\begin{array}{*{20}{c}}
	{\Delta {v^c}}\\
	{\Delta {v^s}}
	\end{array}} \right\} =  - \dfrac{{{G_0}R_B^2}}{4}\left[ {\begin{array}{*{20}{c}}
	1&{\alpha {p_f}}\\
	{\alpha {p_f}}&{{\alpha ^2}}
	\end{array}} \right]\left\{ {\begin{array}{*{20}{c}}
	{{I_c}}\\
	{{I_s}}
	\end{array}} \right\}\\
\;\;\;\;\;\;\;\;\;\;\;\;\;\;\;\;\;\;\;\;\;\;+ \dfrac{{{p_0}{G_0}{R_B}}}{2}\left[ {\begin{array}{*{20}{c}}
	{{p_f}}&{\dfrac{1}{\alpha }}\\
	\alpha &{{p_f}}
	\end{array}} \right]\left\{ {\begin{array}{*{20}{c}}
	{{v_c}}\\
	{{v_s}}
	\end{array}} \right\},
\end{array}
\end{equation}
where $v_c=(v_u+v_d)/2$ and $v_s=\alpha(v_u-v_d)/2$ are charge and spin voltages applied at the external contact, $G_0$ is the contact conductance per unit length, and $p_f$ is the contact polarization with $p_f=0$ indicating a normal metal contact and $p_f\neq 0$ indicating a ferromagnetic contact. The derivation of the model starting from the Boltzmann transport equation is shown in Section \ref{sec_semi} with clearly stated assumptions streamlined with subheadings.

Modified distributed circuit models for charge and spin are shown in Fig. \ref{3}(b) and (c), which are based on Eqs. \eqref{charge_TL_cont}-\eqref{spin_TL_cont} respectively. The presence of a contact with conductance $G_0$ adds series resistances $R_{cont}^c=\dfrac{G_0R_B^2}{4}$ and $R_{cont}^s=\dfrac{\alpha^2 G_0R_B^2}{4}$ in the charge and spin models as shown in Fig. \ref{3}. This effect exists even if the channel is NM. The presence of the external contact also modulates the dependent sources in Eqs. \eqref{dep_Vm} and \eqref{dep_Vn} as
\begin{subequations}
	\begin{equation}
	V_m '=\eta_c V_s+\dfrac{G_0 R_B}{2}\left(\dfrac{v_s}{\alpha}+p_f v_c\right),
	\end{equation}
	\begin{equation}
	V_n '=\eta_s V_c+r_m I_{em}+\dfrac{G_0 R_B}{2}\left(\alpha v_c+p_f v_s\right),
	\end{equation}
\end{subequations}
with $V_a$ and $V_b$ representing the voltage drop across $R_{cont}^c$ and $R_{cont}^s$ in charge and spin models in Fig. \ref{3} respectively. Note that the additional terms are proportional to $G_0$ and negligible for potentiometric contacts where $G_0$ is very low.

\subsection{Special Case: Normal Metals ($p_0=0$)}

We consider a special case of Eqs. \eqref{charge_TL} and \eqref{spin_TL} for a normal metal (NM) channel i.e. $p_0=0$. For NM channel, charge and spin model decouples to well-known models as described below.

\subsubsection{Charge Model: Transport Model for Quantum Wires}

For NM channels ($p_0=0$), the charge model in Eq. \eqref{charge_TL} reduces to the well-known transmission line model for charge that has been previously used to analyze transport in quantum wires, given by
\begin{equation}
\label{Qwire}
\begin{aligned}
&\left(\dfrac{1}{C_E}+\dfrac{1}{C_Q}\right)^{-1}\;\dfrac{\partial}{{\partial t}}{V_c} =  - \dfrac{\partial}{{\partial x}}I_c,\\
&\left(L_K+L_M\right)\;\dfrac{\partial}{{\partial t}}I_c + R_c\,I_c =  - \dfrac{\partial}{\partial x}{V_c}.
\end{aligned}
\end{equation}
The model was first derived from Luttinger liquid theory \cite{Burke_TNANO_2002, Burke_TNANO_2003} and then from Boltzmann transport equation with one electrochemical potential \cite{Sayeef_IEEE_2005}. In the quantum wire limit $L_K\gg L_M$ and $C_Q\ll C_E$ while in the classical transmission line limit $L_K\ll L_M$ and $C_Q\gg C_E$ \cite{Burke_TNANO_2002,Sayeef_IEEE_2005}.

In steady-state ($\partial/\partial t \rightarrow 0$), we get the diffusion equation for charge from Eq. \eqref{Qwire}, given by
\begin{equation}
\dfrac{d^2}{dx^2}V_c=0.
\end{equation}

\subsubsection{Spin Model: Valet-Fert Equation}

For NM channels ($p_0=0$), the spin model in Eq. \eqref{spin_TL} becomes a time-dependent spin-diffusion equation, given by
\begin{equation}
\label{time_Valet_Fert}
\begin{aligned}
&\dfrac{C_Q}{\alpha^2}\dfrac{\partial}{{\partial t}}{V_s} + G_{sh}{V_s} =  - \dfrac{\partial}{{\partial x}}{I_s},\\
&\alpha^2{L_K}\dfrac{\partial}{{\partial t}}{I_s} + {R_s}{I_s} =  - \dfrac{\partial}{{\partial x}}{V_s},
\end{aligned}
\end{equation}
which in steady-state ($\partial/\partial t \rightarrow 0$), becomes the well-known Valet-Fert equation \cite{Valet_Fert_1993}, given by
\begin{equation}
\dfrac{d^2}{dx^2}V_s=\dfrac{V_s}{\lambda_{sf}^2},
\end{equation}
with the spin diffusion length given by
\begin{equation}
\lambda_{sf}=\dfrac{1}{\sqrt{R_s G_{sh}}}=\dfrac{\sqrt{\lambda_0 \lambda_s}}{2}.
\end{equation}
Spin model similar to Eq. \eqref{time_Valet_Fert} has been discussed previously based on Luttinger liquid theory \cite{Burke_TNANO_2002}, however, the model did not take into account the spin relaxation processes in the channel (the shunt conductance $G_{sh}$).

\section{Steady-State Transport Results}
\label{sec_Steady}
In this section, we discuss several established steady-state results on charge-spin interconversion in the potentiometric limit. We start from the steady-state ($\partial/\partial t \rightarrow 0$) form of the transmission line model with external contact in Eqs. \eqref{charge_TL_cont} and \eqref{spin_TL_cont}, given by
\begin{subequations}
	\label{steady_state_TL}
	\begin{alignat}{4}
	&\dfrac{d}{{d x}}I_c = i^c,\\
	&\dfrac{d}{d x}{V_c} = - R_c\,I_c + {p_0}{\eta_c}{V_s} + \Delta v^c,\\
	&\dfrac{d}{{d x}}{I_s} =  -G_{sh}{V_s} + p_0 \gamma_s I_c + i^s,\\
	\text{and}\;\;&\dfrac{d}{{d x}}{V_s} =  - {R_s}{I_s} + p_0 \eta_s {V_c} + \Delta v^s.
	\end{alignat}
\end{subequations}
Under the steady-state condition ($\partial/\partial t \rightarrow 0$), the capacitors and inductors in Fig. \ref{3} become open and short circuits respectively. The steady-state form in Eq. \eqref{steady_state_TL} is equivalent to our prior time-independent semiclassical equations with four electrochemical potentials \cite{Sayed_SciRep_2016} and all our previous results can be reproduced using Eq. \eqref{steady_state_TL}.

We first derive a resistance matrix for a three terminal setup (two charge and one spin) with potentiometric contacts (see Fig. \ref{4}) and assuming reflection with spin-flip to be the dominant scattering mechanism in the channel. We present dc simulation results on charge-spin interconversion in the SML channel using the full model (Eqs. \eqref{charge_TL_cont} and \eqref{spin_TL_cont}) in SPICE and compare with the results from the resistance matrix. We then derive a simple expression for a parameter that has been widely used to quantify inverse Rashba-Edelstein effect (IREE) in 2D channels. We compare the expression with available experiments on diverse materials as well as dc SPICE simulation results using the full model.

\subsection{Resistance Matrix for Potentiometric Setup}

\begin{figure}
	\includegraphics[width=0.49 \textwidth]{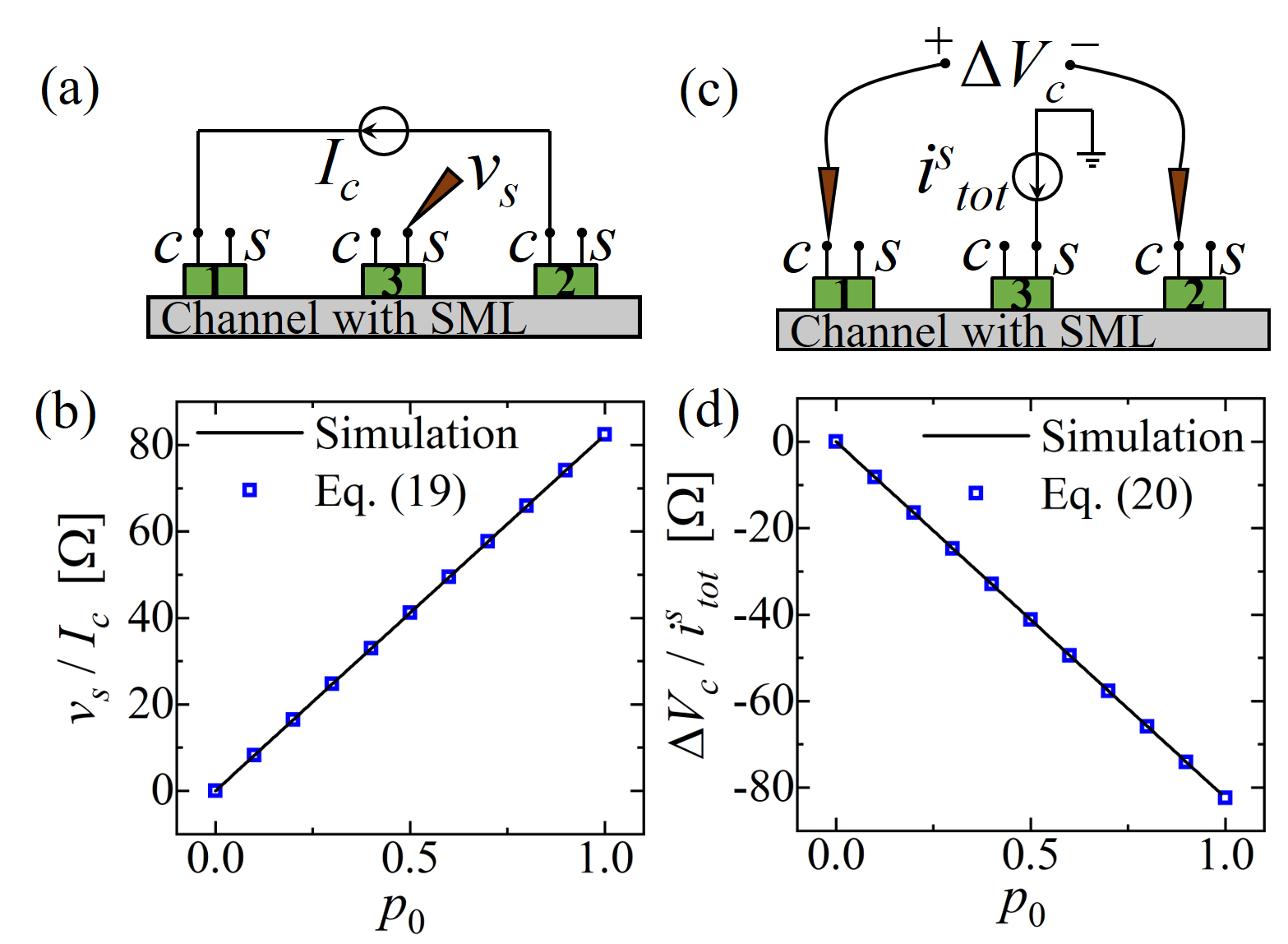}
	\caption{(a) Setup to observe charge current $I_c$ induced spin voltage $v_s$ in the SML channel. Charge terminal of contact 3 is kept open and spin terminals of contacts 1 and 2 are grounded. (b) $v_s/I_c$ vs. $p_0$ for $i^s_{tot}=0$ from SPICE simulation and comparison with Eq. \eqref{Hong_Eqn}. (c) Setup to observe spin current $i^s_{tot}$ induced charge voltage difference $\Delta V_c$ across the SML channel. Charge terminal of contact 3 is kept open and spin terminals of contacts 1 and 3 are grounded. (d) $\Delta V_c/i^s_{tot}$ vs. $p_0$ for $I_c=0$ from SPICE simulation and comparison with Eq. \eqref{Hong_Inv}. The SPICE setup is shown in Fig. \ref{3_7} of Appendix \ref{App_Hong}. Parameters: $\alpha=2/\pi$, $G_0\approx0.05G_B$, $M+N=100$, and scattering rate per unit mode $=0.04$ per lattice points. We assumed reflection with spin-flip to be the dominant scattering process in the channel.}\label{4}
\end{figure}

We consider a structure with three NM contacts ($p_f=0$) on top of a SML channel, as shown in Fig. \ref{4}. Contacts 1 and 2 are the charge terminals and contact 3 is the spin terminal with no charge current flowing through it (i.e. $i^c=0$). We start from Eq. \eqref{steady_state_TL} and make two assumptions to drive the resistance matrix: (i) the contacts are potentiometric i.e. the contact conductance per unit length $G_0$ is very low such that the following condition is satisfied
\begin{equation}
\label{pot_cond}
\dfrac{1}{\lambda},\dfrac{1}{\lambda_0}, \dfrac{1}{\lambda_r}\gg \dfrac{G_0 R_B}{4},
\end{equation}
and (ii) the reflection with spin-flip is the dominant scattering mechanism in the channel. The details of the derivation are given in Appendix \ref{App_Rmat}.

The resistance matrix is given by
\begin{equation}
\label{R_Mat}
\left\{ {\begin{array}{*{20}{c}}
	\Delta V_c\\
	{{v_{s}}}
	\end{array}} \right\} = \left[ {\begin{array}{*{20}{c}}
	{\dfrac{R_B L}{{\lambda }}}+\dfrac{2}{G_0''}&{ - \dfrac{{\alpha {p_0} R_B}}{{2}}}\\
	{\dfrac{{\alpha {p_0} R_B}}{{2}}}&{ \dfrac{{{\alpha ^2}}}{{{G_0}^\prime }}}
	\end{array}} \right]\left\{ {\begin{array}{*{20}{c}}
	{{I_c}}\\
	{i_{tot}^s}
	\end{array}} \right\},
\end{equation}
where $\Delta V_c$ is the charge voltage difference between contacts 1 and 2, $I_c$ is the charge current flowing in the channel with length $L$, $v_s$ is the spin voltage at contact 3, $i^s_{tot}=i^sL$ is the spin current at contact 3, contacts 1 and 2 have equal conductance $G_0''$, and $G_0'=G_0 L$ is the contact conductance of contact 3. Eq. \eqref{R_Mat} is the similar to that previously reported in Ref. \cite{Hong_SciRep_2016} with corrections in the diagonal components. The diagonal components $(1,1)$ and $(2,2)$ in the matrix represent the charge resistance between contacts 1 and 2 and spin resistance at contact 3 respectively. 

Note that Eq. \eqref{R_Mat} is derived under the assumption that reflection with spin-flip is the dominant scattering mechanism. In the presence of other scattering mechanisms, the basic structure of Eq. \eqref{R_Mat} remains the same, however, additional factors related to scattering rates multiply each of the components in the matrix (see Appendix \ref{App_Rmat}).

\subsection{Direct Effects: Charge Current Induced Spin Voltage}
For a charge current $I_c$ flowing through the SML channel, the open circuit spin voltage $v_s$ at the contact 3 (i.e. $i^s_{tot}=0$) can be derived from Eq. \eqref{R_Mat} as
\begin{equation}
\label{Hong_Eqn}
v_s = \dfrac{\alpha p_0}{2 G_B}I_c.
\end{equation}
Eq. \eqref{Hong_Eqn} was originally proposed in Ref. \cite{Hong_PRB_2012} which was later confirmed by a number of experiments \cite{JonkerNatNano2014, KLWangNanoLett2014, DasNanoLett2015, SamarthPRB2015, YPChenSciRep2015, Samarth_PRB_2015, YoichiPRB2016, Koo_New} on different TI materials using potentiometric ferromagnetic contact. Note that the spin voltage measured using a contact with higher conductance will be lower than Eq. \eqref{Hong_Eqn} due to the current shunting effect in the contact \cite{Sayed_SciRep_2016}. The full model in Eqs. \eqref{charge_TL_cont} and \eqref{spin_TL_cont} takes into account such effect related to contact conductances.

We have simulated the structure in Fig. \ref{4}(a) by connecting the full two-component circuit models given in Figs. \ref{1} and \ref{3} in a distributed manner using the standard circuit rules. The details of the simulation are discussed in Appendix \ref{App_Hong}. The simulation results of $v_s / I_c$ as a function of $p_0$ is shown in Fig. \ref{4}(b), which is in good agreement with Eq. \eqref{Hong_Eqn}.

\subsection{Inverse Effects: Spin Current Induced Charge Voltage}

The reciprocal effect of Eq. \eqref{Hong_Eqn} is the spin current $i_s$ induced open circuit charge voltage difference $\Delta V_c$ across the SML channel \cite{Hong_SciRep_2016}. For $I_c=0$ we have from Eq. \eqref{R_Mat}
\begin{equation}
\label{Hong_Inv}
\Delta V_c = - \dfrac{\alpha p_0}{2 G_B}i^s_{tot}.
\end{equation}
The reciprocal relation between Eqs. \eqref{Hong_Eqn} and \eqref{Hong_Inv} including the negative sign has been observed experimentally \cite{SamarthPRB2015, Pham_NanoLett_2016, Koo_New}. Note that Eq. \eqref{Hong_Inv} is exact for a potentiometric contact. For a contact with higher conductance, $\Delta V_c$ will be lower than Eq. \eqref{Hong_Inv} due to current shunting by the same amount as that for the direct effect (in Eq. \eqref{Hong_Eqn}) \cite{Sayed_SciRep_2016}, which can be analyzed using the full model in Eqs. \eqref{charge_TL_cont} and \eqref{spin_TL_cont}. We have simulated the structure shown in Fig. \ref{4}(c) using the full two-component models in Figs. \ref{1} and \ref{3}. The simulation results of $\Delta V_c / i^s_{tot}$ as a function of $p_0$ is shown in Fig. \ref{4}(d), which show good agreement with Eq. \eqref{Hong_Inv}. The details of the simulation are discussed in Appendix \ref{App_Hong}.

\begin{table}
	\begin{center}
		\caption{Estimated Material Parameters.}
		\label{table_IREE}
		\begin{tabular}{||c | c | c | c | c ||} 
			\hline
			\label{IREE_table}
			Material & $\lambda$ [nm] & $p_0$ & $\lambda_{IREE}$ [nm] & $\lambda_{IREE}$ [nm]\\&&&(Eq. \eqref{IREE_length})&(measured)\\ [0.5ex] 
			\hline\hline
			Ag$|$Bi & 22.6$^\dag$  & 0.05$^{\dag\dag}$ & 0.36 & 0.3 \cite{Fert_NatComm_2016}\\
			\hline
			Cu$|$Bi & 0.88$^\dag$ & 0.05$^{\dag\dag}$ & 0.014 & 0.009 \cite{IssaCuBi2016}\\
			\hline
			LAO$|$STO & 180.8$^\dag$ & 0.0616$^{\dag\dag}$ & 3.55 & 6.4 \cite{FertLAOSTO2016}\\
			\hline
			Bi$_2$Se$_3$ & 3.2$^\dag$ & 0.025$^{\ddagger}$ & 0.026 & 0.035 \cite{SmarthPRL2016}\\
			\hline
		\end{tabular}
	\end{center}
	\begin{flushleft}
		$^\dag${\footnotesize Estimated from the measured sheet resistance of the samples.}\\$^{\dag\dag}${\footnotesize Estimated from the Rashba coupling coefficient and Fermi velocity of the materials.\\$^{\ddagger}${\footnotesize Estimated from spin-pumping induced voltage and Eq. \eqref{Hong_Inv}.}\\The estimations are discussed in detail in Appendix \ref{App_IREE}.}
	\end{flushleft}
\end{table}

\subsection{Inverse Rashba-Edelstein Effect (IREE) Length}
The phenomena described by Eq. \eqref{Hong_Inv} is often known as the inverse Rashba Edelstein effect (IREE) for 2D channels. IREE is often quantified with a parameter called IREE length defined as 
\begin{equation}
\label{defIREE}
\lambda_{IREE}=\dfrac{J_c}{J_s},
\end{equation}
where $J_c$ is the longitudinal short circuit charge current density in A/m induced by the injected transverse spin current density $J_s$ in A/m$^2$.

We derive a simple expression for IREE length starting from the first row of Eq. \eqref{R_Mat} for short circuit condition between contacts 1 and 2 (i.e. $v_{c1}=v_{c2}$) and assuming large channel resistance compared to contact resistance (i.e. $L/(G_B\lambda)\gg2/G_0''$). The expression is given by
\begin{equation}
\label{IREE_length1}
\lambda_{IREE} = \dfrac{\alpha p_0 \lambda}{2}.
\end{equation}
The derivation is given in Appendix \ref{App_Rmat}. Note that both $p_0$ and $\lambda$ are two completely independent parameters.

\begin{figure}
	\includegraphics[width=0.4 \textwidth]{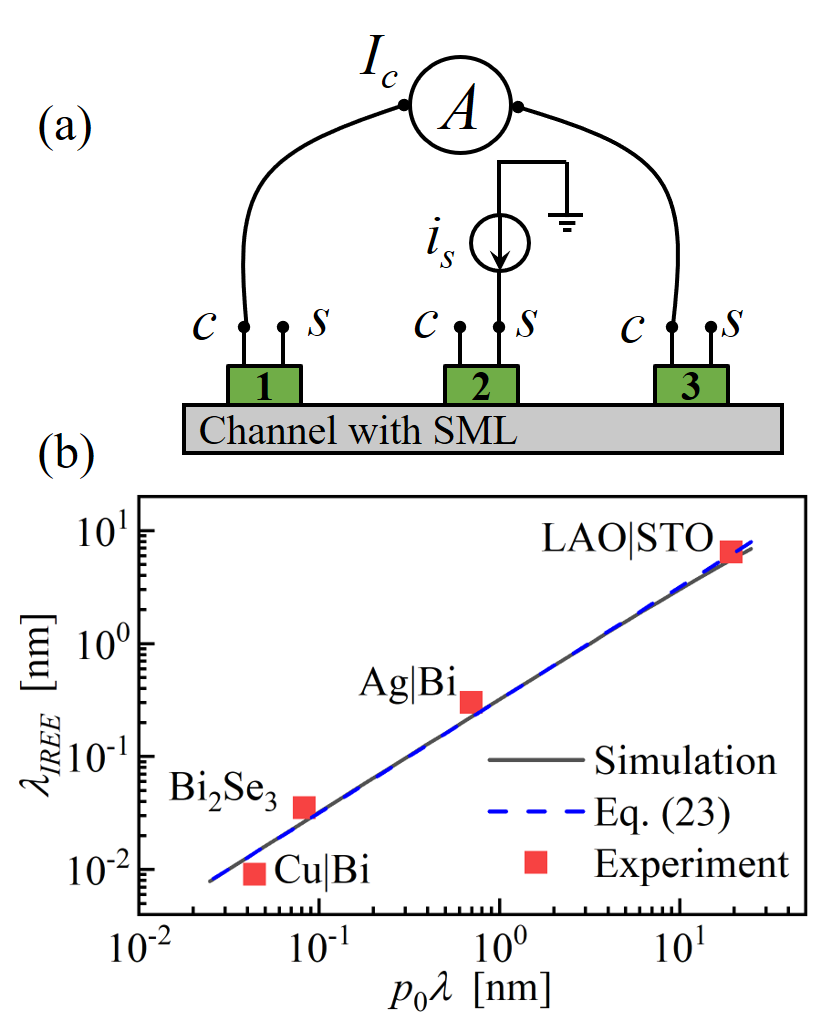}
	\caption{ (a) Setup similar to that in Fig. \ref{4}(c) but short circuit charge current between contacts 1 and 2 is being observed. (b) Inverse Rashba Edelstein effect (IREE) length ($\lambda_{IREE}$) vs. $p_0\lambda$ from SPICE simulation and comparison to Eq. \eqref{IREE_length} and experiments on different interfaces with Rashba SOC: Ag$|$Bi \cite{Fert_NatComm_2016}, Cu$|$Bi \cite{IssaCuBi2016}, LaAlO$_3$$|$SrTiO$_3$ (LAO$|$STO) \cite{FertLAOSTO2016}, and Bi$_2$Se$_3$ \cite{SmarthPRL2016}. The back scattering length $\lambda$ is estimated from measured sheet resistance or resistivity. The degree of spin-momentum locking $p_0$ is estimated from the Rashba coupling coefficient and the Fermi velocity using Eq. \eqref{RashbaSML}. The details of estimations are given in Appendix \ref{App_Hong}. SPICE simulation parameters: $\alpha=2/\pi$, $G_0\approx0.05G_B$, and $M+N=100$. We assumed reflection with spin-flip to be the dominant scattering process in the channel.}\label{5}
\end{figure}

Here, $\alpha$ is an angular averaging factor that can vary between 0 and 1 depending on the angular variation of the spin polarization of the eigenstates and details of the scattering mechanism. We assume that the distribution is such that the angle between $z$-axis and the spin polarization of the eigenstates with particular group velocity ($+$ or $-$) vary between $-\pi/2$ to $+\pi/2$, which yields $\alpha=2/\pi$. Thus, from Eq. \eqref{IREE_length1} we have the following expression
\begin{equation}
\label{IREE_length}
\lambda_{IREE}=\dfrac{p_0\lambda}{\pi},
\end{equation}
which has been previously reported in Ref. \cite{Sayed_SciRep_2016}. We have simulated the setup in Fig. \ref{5}(a) in order to estimate $\lambda_{IREE}$ of diverse ranges of $p_0$ and $\lambda$. The setup in Fig. \ref{5}(a) is same as that in Fig. \ref{4}(c), except we observe the short circuit charge current $I_c$ between the charge terminals of contacts 1 and 2 induced by the injected spin current $i_s$ through the spin terminal of contact 3.

We have compared the simulation results with Eq. \eqref{IREE_length} as well as available measurements from spin-pumping and lateral spin valve experiments on different Rashba interfaces: Ag$|$Bi \cite{Fert_NatComm_2016}, Cu$|$Bi \cite{IssaCuBi2016}, LaAlO$_3$$|$SrTiO$_3$ (LAO$|$STO) \cite{FertLAOSTO2016}, and Bi$_2$Se$_3$ \cite{SmarthPRL2016}. The comparison is shown in Fig. \ref{5}(b). We have estimated $\lambda$ from the reported sheet resistance or resistivity of the samples. $p_0$ for Ag$|$Bi, Cu$|$Bi, and LAO$|$STO are estimated using the Rashba coupling coefficient ($v_0$) and the Fermi velocity ($v_F$) quoted in the literature, using the following expression:
\begin{equation}
\label{RashbaSML}
p_0 = \dfrac{{{v_0}}}{{\sqrt {v_0^2 + v_F^2}}}.
\end{equation}
For Bi$_2$Se$_3$, we have estimated $p_0$ from spin-pumping induced inverse voltage and Eq. \eqref{Hong_Inv}. Note that $p_0$ estimated for the sample in Ref. \cite{SmarthPRL2016} is much lower than previous reports \cite{JonkerNatNano2014, KLWangNanoLett2014, DasNanoLett2015, SamarthPRB2015, YPChenSciRep2015, Samarth_PRB_2015, YoichiPRB2016}, which may be due to the presence of large number of parallel channels. We expect higher $\lambda_{IREE}$ for Bi$_2$Se$_3$ samples with higher $p_0$.

The derivation of Eq. \eqref{RashbaSML} from the Rashba Hamiltonian is shown in Appendix \ref{App_IREE}. The estimations are summarized in Table \ref{table_IREE} and the details are given in Appendix \ref{App_IREE}. These two independent estimations of $p_0$ and $\lambda$ when applied to Eq. \eqref{IREE_length}, agrees very well with experimentally reported $\lambda_{IREE}$, as shown in Fig. \ref{5} and Table \ref{table_IREE}.

\section{Time-Dependent Transport Results}
In this section, we use the full time-dependent model in Eqs. \eqref{charge_TL} and \eqref{spin_TL} to discuss a well-known phenomenon called the spin-charge separation. The spin-charge separation in the presence of spin-orbit coupling is a subject that has been controversial in the past. Our model shows that the charge and spin propagates with two distinct velocities which persist even in the materials with spin-orbit coupling exhibiting spin-momentum locking ($p_0\neq 0)$.  However, we show that the lower velocity signal is purely spin while the higher velocity signal is largely charge with an additional spin component proportional to $p_0$.

\subsection{Velocities of Charge and Spin Signals}
The velocities for charge and spin signals can be derived by finding the eigenvalues of Eqs. \eqref{charge_TL} and \eqref{spin_TL} assuming the low loss limit and constant coefficients. In addition, we find the corresponding eigenvectors as well to analyze the coupling between spin and charge in the channel due to SML. The details of the derivation are given in Appendix \ref{AppA}.

The lower velocity eigenvalue is given by
\begin{equation}
\label{spin_vel}
v_{g,s}=\pm\dfrac{1}{\sqrt{L_K C_Q}}= \pm\langle {v}_x\rangle,
\end{equation}
which is determined by quantum capacitance $C_Q$ and kinetic inductance $L_K$, resulting in thermally averaged electron velocity $\langle{v}_x\rangle$. The corresponding eigenvector is given by
\begin{equation}
\label{spin_eig}
\left\{ {\begin{array}{*{20}{c}}
	{{V_c}}\\
	{{I_c}}
	\end{array}} \right\} = \left\{ {\begin{array}{*{20}{c}}
	0\\
	0
	\end{array}} \right\}, \text{ and } \left\{ {\begin{array}{*{20}{c}}
	{{V_s}}\\
	{{I_s}}
	\end{array}} \right\} = \left\{ {\begin{array}{*{20}{c}}
	{ \pm \alpha^2 \sqrt {\dfrac{{{L_K}}}{{{C_Q}}}} }\\
	1
	\end{array}} \right\},
\end{equation}
which shows that the lower velocity signal is purely spin and no charge accompanies the signal even in channels with SML i.e. $p_0 \neq 0$.

The higher velocity eigenvalue is given by
\begin{equation}
\label{charge_vel}
v_{g,c}=\pm\dfrac{1}{\sqrt{L_{eff} C_{eff}}},
\end{equation}
where $C_{eff}$ is a series combination of $C_E$ and $C_Q$ and $L_{eff}$ is a series combination of $L_M$ and $L_K$. The corresponding eigenvector is given by
\begin{equation}
\label{charge_eig}
\begin{array}{*{20}{c}}
\left\{ {\begin{array}{*{20}{c}}
	{{V_c}}\\
	{{I_c}}
	\end{array}} \right\} = \left\{ {\begin{array}{*{20}{c}}
	{ \pm \sqrt {\dfrac{{{L_{eff}}}}{{{C_{eff}}}}} }\\
	1
	\end{array}} \right\}, \text{ and}\\
\left\{ {\begin{array}{*{20}{c}}
	{{V_s}}\\
	{{I_s}}
	\end{array}} \right\} = \dfrac{{{p_0}}}{{v_{g,c}^2 - v_{g,s}^2}}\left\{ {\begin{array}{*{20}{c}}
	{ - \alpha^2 {g_m}\dfrac{{{L_M}}}{{{C_Q}}}v_{g,c}^2 + {r_m}v_{g,s}^2}\\
	{ \pm {{{v_{g,c}}{v_{g,s}^2}}}{{{C_Q}}}\left( { - {g_m}\dfrac{{{L_M}}}{{{C_Q}}} + \dfrac{r_m}{\alpha^2}} \right)}
	\end{array}} \right\},
\end{array}
\end{equation}
which shows that the higher velocity signal is largely charge which will be accompanying an additional spin signal proportional to $p_0$, which has not been discussed before. This additional spin component vanishes in a NM channel where there is no SML (i.e. $p_0=0$) and the signal is purely charge. Further evaluation of this high velocity spin component we leave as future work.

The quantum capacitance $C_Q$ is proportional to the total number of modes $(M+N)$ in the channel (see Eq. \eqref{cq}) while the kinetic inductance $L_K$ is inversely proportional to $M+N$ (see Eq. \eqref{lk}). $M+N$ is proportional to the channel width (for 2D) or cross-sectional area (for 3D) \cite{Datta_LNE_2012}.

For a conductor with very large cross-section, we may have $C_E\ll C_Q$ and $L_M\gg L_K$ which is the standard transmission line limit. In this limit, the velocity in Eq. \eqref{charge_vel} becomes
\begin{equation*}
c=\dfrac{1}{\sqrt{L_MC_E}},
\end{equation*}
which is the velocity predicted by standard transmission line theory and can be as high as the speed of light.

For a conductor with very small cross-section like quantum wires, we may have $C_E\gg C_Q$ and $L_M\ll L_K$. In this limit, velocity in Eq. \eqref{charge_vel} becomes the thermally averaged electron velocity, given by
\begin{equation*}
\langle{v}_x\rangle=\dfrac{1}{\sqrt{L_KC_Q}},
\end{equation*}
which is same as Eq. \eqref{spin_vel}. Note that the two velocity eigenvalues are equal at this limit.


\subsection{Spin-Charge Separation}

From Eqs. \eqref{spin_vel} and \eqref{charge_vel} we have
\begin{equation}
\dfrac{v_{g,s}}{v_{g,c}}=\sqrt{\dfrac{1+\delta \dfrac{C_Q}{C_E}}{1+ \dfrac{C_Q}{C_E}}},
\end{equation}
where $\delta = \left(L_M C_E\right)/\left(L_K C_Q\right) = \left(\langle v_x\rangle/c\right)^2$ (see Eq. 6 of Ref. \cite{Sayeef_IEEE_2005}) which is usually much less than one, making the spin signal slower than the charge signal, given by
\begin{equation}
\label{SCS}
v_{g,s} < v_{g,c}.
\end{equation}
This results in spin-charge separation which is well-established for channels without SML (i.e. $p_0=0$) from Luttinger liquid theory (see for example, Refs. \cite{HalperinJAP2007, PoliniPRL2007, SchroerPRL2014, Burke_TNANO_2002}, and references therein). Electrons' charge excites $C_E$ and $L_M$ hence the charge signal velocity is determined by $C_{eff}$ and $L_{eff}$ given by Eq. \eqref{charge_vel}. However, pure spin signal do not excite $C_E$ and $L_M$, hence its velocity is determined by $C_Q$ and $L_K$ only given by Eq. \eqref{spin_vel}. Similar arguments have been made in the past \cite{Burke_TNANO_2002, Burke_TNANO_2003} in the context of carbon nanotubes without SOC.

Note that the argument in Eq. \eqref{SCS} is independent of $p_0$ (see Appendix \ref{AppA}), which indicates that the spin-charge separation persists even in channels with SOC exhibiting SML (i.e. $p_0\neq 0$). Similar arguments have been discussed previously by considering SOC \cite{BalseiroPRL2002, CalzonaPRB2015, Stauber_PRB_2013} although there exists argument that the presence of SOC may destroy the spin-charge separation \cite{BarnesPRL2000}.

In SML channels, an additional spin signal proportional to $p_0$ accompanies the charge signal at the same velocity as the charge ($v_{g,c}$) as seen from Eq. \eqref{charge_eig}. This additional spin component is induced by the instantaneous voltage drop across $L_M$ and the instantaneous current through $C_E$ of the channel. However, the low velocity signal (see Eq. \eqref{spin_vel}) remains purely spin since the spin signal do not excite $L_M$ and $C_E$ \cite{Burke_TNANO_2002,Sayeef_IEEE_2005} to induce a similar accompanying charge component.

\section{Transmission Line Model from Boltzmann Formalism}
\label{sec_semi}

In this section, we derive the transmission line model in Eqs. \eqref{charge_TL_cont} and \eqref{spin_TL_cont} starting from the time-dependent Boltzmann transport equation under several clearly stated assumptions, which allow us to obtain the simple expressions for the model parameters stated in Eq. \eqref{params}. Several of our predictions for steady-state \cite{Sayed_SciRep_2016} have already received experimental support \cite{Pham_APL_2016,Pham_NanoLett_2016,Koo_New} suggesting that the assumptions are reasonable, but they could be revisited as the field evolves.

\subsection{Boltzmann Transport Equation}

We assume a structure where the spatial variations and the applied fields are along $\hat{x}$-direction. The time-dependent Boltzmann transport equation is given by
\begin{equation}
\label{BTE_1D}
\begin{aligned}
\frac{\partial f}{{\partial t}} + {v}_x \frac{\partial f}{{\partial x}} + {F}_x \frac{\partial f}{{\partial p_x}} = \displaystyle \sum_{\vec{p'}, s '} S(\vec{p},s\leftrightarrow\vec{p'},s') \left(f-f'\right),
\end{aligned}
\end{equation}
where we have assumed elastic scattering so that the scattering rates are same in both directions i.e. 
\begin{equation*}
S(\vec{p},s\rightarrow\vec{p'},s')=S(\vec{p'},s'\rightarrow\vec{p},s)
\equiv S(\vec{p},s\leftrightarrow\vec{p'},s').
\end{equation*}

Here, $f\equiv f(x,t,\vec{p},s)$ is the occupation factor of a state for a particular position $x$, time $t$, momentum $\vec{p}$ and spin index $s=\pm1$, $f'\equiv f(x,t,\vec{p'},s')$ with momentum $\vec{p'}$ and spin index $s'=\pm1$, ${v}_x=\partial E / \partial p_x$ is the $x$-component of the group velocity, ${F}_x=-\partial E/\partial x$ is the force on electrons along $\hat{x}$-direction, and $E$ is the total energy. Note that the spin index $+1$ and $-1$ correspond to up ($U$) and down ($D$) spin polarized states for a particular $\vec{p}$.

\subsection{Occupation Factor}
We write the occupation factor $f$ in terms of an electrochemical potential $\mu\equiv\mu(x,t,\vec{p},{s})$, in the form
\begin{equation}
\label{occup_fact}
f(x,t,\vec{p},{s})=\dfrac{1}{1  + \exp\left(\dfrac{E(x,t,\vec{p},{s})-\mu(x,t,\vec{p},s)}{k_BT} \right)},
\end{equation}
where $k_B$ is the Boltzmann constant, $T$ is the temperature. Note that 

\subsection{Linearization}
We apply a variable transform $\xi= E-\mu$ on the left hand side of Eq. \eqref{BTE_1D}. On the right hand side of Eq. \eqref{BTE_1D}, we expand both $f$ and $f'$ into Taylor series around
\begin{equation}
\label{eq_occup_fact}
f_0 = \dfrac{1}{1+\exp\left(\dfrac{(E(x,\vec{p},{s})-\mu_0)}{k_BT}\right)},
\end{equation}
with constant electrochemical potential $\mu_0$ and apply linear response approximation. Thus Eq. \eqref{BTE_1D} can be written as (see Appendix \ref{AppB} for details of the derivation)
\begin{equation}
\label{semi_1D_full}
\begin{array}{l}
\left(-\dfrac{\partial f_0}{\partial E}\right)\left(\dfrac{\partial \mu}{{\partial t}} - \dfrac{\partial E }{{\partial t}} + {v}_x \dfrac{\partial \mu}{{\partial x}} + {F}_x \dfrac{\partial \mu}{{\partial p_x}}\right) \\\quad\quad = -\displaystyle \sum_{\vec{p'},s'}S(\vec{p},s\leftrightarrow\vec{p'},s') \left(-\dfrac{\partial f_0}{\partial E}\right){\left(\mu - \mu ' \right)}.
\end{array}
\end{equation}

We assume that there are no internal fields in the present discussion. Hence, $F_x$ comes from the applied voltage and the term $F_x(\partial \mu /\partial p_x)$ depends on the higher order of the applied voltage, which can be neglected in the linear response regime \cite{Datta_LNE_2012}. Thus Eq. \eqref{semi_1D_full} is given by
\begin{equation}
\label{semi_1D}
\begin{array}{l}
\left(-\dfrac{\partial f_0}{\partial E}\right)\left(\dfrac{\partial \mu}{{\partial t}} - \dfrac{\partial E }{{\partial t}} + {v}_x \dfrac{\partial \mu}{{\partial x}} \right) \\\quad\quad= -\displaystyle \sum_{\vec{p'},s'}S(\vec{p},s\leftrightarrow\vec{p'},s') \left(-\dfrac{\partial f_0}{\partial E}\right){\left(\mu - \mu ' \right)}.
\end{array}
\end{equation}
The term $\partial E / \partial t$ can be evaluated from the dispersion relation of a given Hamiltonian in the semiclassical approximation as discussed below.

\subsection{Dispersion Relation}
We start from the following Rashba Hamiltonian
\begin{equation}
\label{Rashba}
\mathcal{H}=\dfrac{|\vec p -q \vec A|^2}{2m}I_{2\times2}-v_0 \left(\vec \sigma \times (\vec p-q\vec A)\right)\cdot\hat{y} + U_E I_{2\times2}.
\end{equation}
Here, $I_{2\times2}$ is a $2\times2$ identity matrix,  $\vec p$ and $\vec{A}$ are the momentum and vector magnetic potential respectively in the $z$-$x$ plane, $\vec{\sigma}$ is the Pauli's matrices, $v_0$ is the Rashba coefficient, $U_E$ is the electrostatic potential, $m$ is the electron mass, and $q$ is the electron charge.

Eigenstates of Eq. \eqref{Rashba} are given by
\begin{equation}
\label{E_eig}
E(\vec{p},{s})=\dfrac{|\vec{p}-q\vec{A}|^2}{2m}-s\, v_0 {|\vec{p}-q\vec{A}|} + U_E,
\end{equation}
with $\vec{p}$ is confined to the $z$-$x$ plane.

We assume that $U_E$ and $\vec{A}$ varies slowly with $x$ and $t$, so that in the semiclassical approximation we have
\begin{equation}
\label{E_rel}
E(x,t,\vec{p},{s})=\dfrac{|\vec{p}-q\vec{A}(x,t)|^2}{2m}-s\, v_0 {|\vec{p}-q\vec{A}(x,t)|} + U_E(x,t).
\end{equation}

Differentiating Eq. \eqref{E_rel} with respect to $t$ yields
\begin{equation}
\label{dE_rel}
\dfrac{\partial E}{\partial t} = \vec{v}\cdot \left(-q\dfrac{\partial \vec{A}}{\partial t}\right) + \dfrac{\partial U_E}{\partial t},
\end{equation}
where $\vec{v}=\nabla_{\vec{p}}E$ (see Appendix \ref{AppC} for the derivation). 

The electrostatic potential $U_E$ and the vector magnetic potential $\vec A$ on the structure of interest can be evaluated from the theory of electromagnetism.

\subsection{From Potentials to Charge and Current}

The electrostatic potential $U_E$ is related to the total charge $Q$ in the channel by the electrostatic capacitance $C_E$ of the structure under consideration, given by
\begin{equation}
\dfrac{U_E}{q}=\dfrac{Q}{C_E}.
\end{equation}

We assume that the charge current $I_c$ flows along $\hat{x}$-direction which is uniform in the channel. The vector magnetic potential $\vec{A}$ is related to $I_c$ by the magnetic inductance $L_M$ of the channel, given by
\begin{equation}
\label{vec_mag_assum}
\vec{A}\equiv \hat{x}A_x = \hat{x} L_MI_c.
\end{equation}

Thus Eq. \eqref{dE_rel} can be written as
\begin{equation}
\label{dE_rel2}
\dfrac{\partial E }{\partial t} = -q v_x L_M \dfrac{\partial I_c}{\partial t} + \dfrac{q}{C_E}\dfrac{\partial Q}{\partial t}.
\end{equation}

We combine Eq. \eqref{semi_1D} with Eq. \eqref{dE_rel2} to get
\begin{equation}
\label{semi_1mu}
\begin{aligned}
\left(\dfrac{\partial f_0}{\partial E}\right)&\left(\frac{\partial \mu}{{\partial t}} + {v}_x \frac{\partial \mu}{{\partial x}} + q v_x L_M \dfrac{\partial I_c}{\partial t} - \dfrac{q}{C_E}\dfrac{\partial Q}{\partial t}\right) \\& = -\displaystyle \sum_{\vec{p'},s'}S(\vec{p},s\leftrightarrow\vec{p'},s') \left(\dfrac{\partial f_0}{\partial E}\right){\left(\mu - \mu ' \right)}.
\end{aligned}
\end{equation}

\subsection{Classification}
\begin{figure}
	\includegraphics[width=0.35 \textwidth]{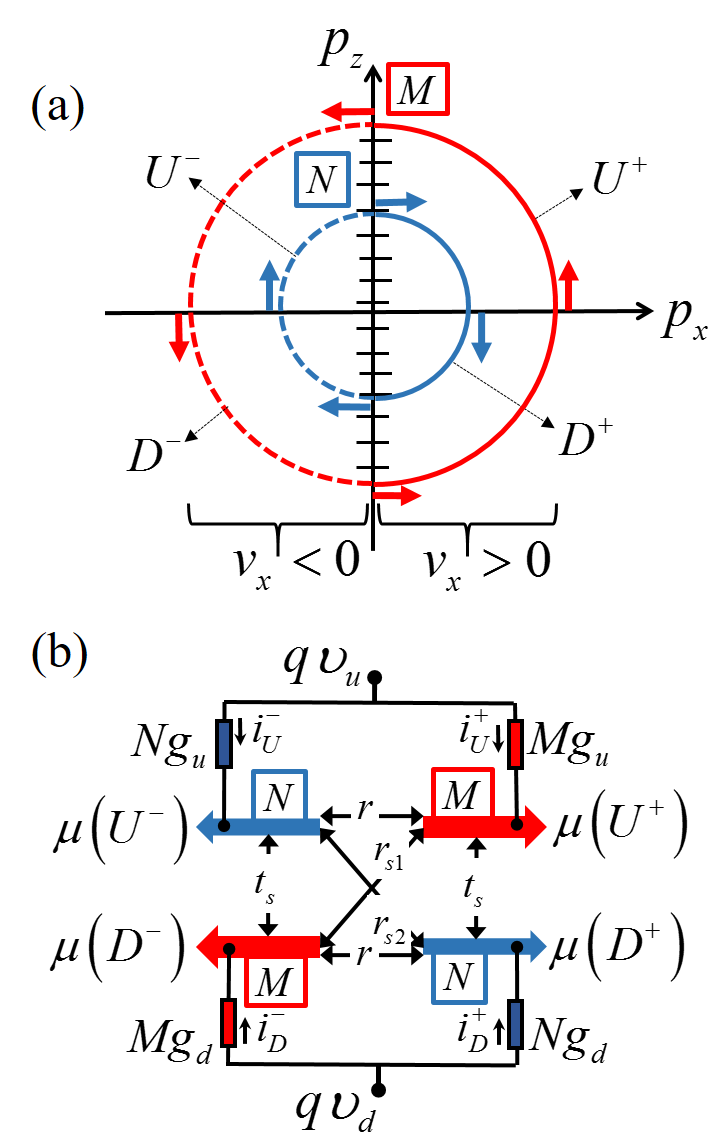}
	\caption{(a) Two Fermi circles of a Rashba channel with spin-momentum locking having opposite spin polarizations at a given energy $E_F$. Left (right) half of each circle represents negative (positive) group velocity. The large circle corresponds to the large number of modes $M$ in the channel and have a net spin polarization along $+\hat{z}$ ($-\hat{z}$) on right (left) side. The small circle corresponds to smaller number of modes $N$ and have net spin polarization along $-\hat{z}$ ($+\hat{z}$) on right (left) side. We classify all electronics states into four groups based on the $z$-component of spin polarization (up ($U$), down ($D$)) and the sign of $x$-component of the group velocity ($+$, $-$). (b) Assuming high scattering among states in each individual groups, we assign four electrochemical potentials for these four groups: $\mu(U^+)$, $\mu(U^-)$, $\mu(D^+)$, and $\mu(D^-)$. External contact is modeled as up ($v_u$) and down ($v_d$) spin voltages applied to up and down states of the channel through up ($g_u$) and down ($g_d$) spin conductances per mode, respectively. Four different currents ($i_U^+$, $i_D^+$, $i_U^-$, and $i_D^-$) enter the four different groups in the channel. The contact can be either normal metal $g_u = g_d$ or ferromagnet $g_u \neq g_d$.}\label{2}
\end{figure}

We classify all $\vec{p},s$ states into four groups based on the sign of $v_x$ ($+$ or $-$) and the spin index $s=\pm1$, given by
\begin{equation}
\label{classification}
\begin{aligned}
\Re:
\begin{cases} 
&U^+\in\{\vec{p},s\,\,\,|\,\,\,v_x>0,\,\,\,s=+1\},\\
&D^-\in\{\vec{p},s\,\,\,|\,\,\,v_x<0,\,\,\,s=-1\},\\
&U^-\in\{\vec{p},s\,\,\,|\,\,\,v_x<0,\,\,\,s=+1\},\text{ and}\\
&D^+\in\{\vec{p},s\,\,\,|\,\,\,v_x>0,\,\,\,s=-1\}.
\end{cases}
\end{aligned}
\end{equation}
where $s=+1$ and $-1$ denote up ($U$) and down ($D$) spins with respect to the spin quantization axis defined by $\hat{y}\times\left(\vec{p}-q\vec{A}\right)$, which is different for each direction of $\vec{p}$.
Such classification can be mapped onto the two Fermi circles of a Rashba channel (see Fig. \ref{2}(a)). The large circle corresponds to $U^+$ and $D^-$ groups which share the same number of modes $n_m(U^+)=n_m(D^-)=M$ satisfying the time-reversal symmetry. Similarly, the small circle corresponds to $U^-$ and $D^+$ groups sharing the same number of modes $n_m(U^-)=n_m(D^+)=N$. Note that the eigenstates belonging to each of the four half Fermi circles in Fig. \ref{2}(a) has an average spin polarization along $\hat{z}$-direction with an averaging factor of $\alpha = 2/\pi$, which we will use later when writing spin currents and voltages.

\subsection{Averaging}
We define the thermal average of a variable $\psi\equiv\psi(\vec{p},s)$ within each of the group $\vec{p},s\in\Re$ as
\begin{equation}
\label{th_avg}
\langle\psi\rangle_{\vec{p},s\in\Re} = \dfrac{\displaystyle\sum_{\vec{p},s\in\Re}\left(-\dfrac{\partial f_0}{\partial E}\right)\psi(\vec{p},s)}{\displaystyle\sum_{\vec{p},s\in\Re}\left(-\dfrac{\partial f_0}{\partial E}\right)}.
\end{equation}

We sum both sides of Eq. \eqref{semi_1mu} over all $\vec p$ states within range $\vec{p},s\in\Re$ in the $z$-$x$ plane as
\begin{equation}
\label{semi_sum}
\begin{aligned}
&\dfrac{D_0(\Re)}{2}\dfrac{\partial \left\langle \mu  \right\rangle }{\partial t} + \dfrac{D_0(\Re)}{2}\dfrac{\partial \langle {v_x\mu} \rangle}{\partial x} \\&+ \dfrac{qD_0(\Re)}{2}\langle v_x\rangle L_M \dfrac{\partial I_c}{\partial t}
- \dfrac{q}{C_E}\dfrac{D_0(\Re)}{2}\dfrac{\partial Q}{\partial t} \\&= -\displaystyle \sum_{\vec{p},s\in\Re}\,\, \displaystyle \sum_{\vec{p'},s'}S(\vec{p},s\leftrightarrow\vec{p'},s') \left(-\dfrac{\partial f_0}{\partial E}\right){\left(\mu - \mu ' \right)}.
\end{aligned}
\end{equation}
Here, $D_0(\Re)$ is the thermally averaged density of states within $\vec{p},s\in\Re$ given by
\begin{equation}
\label{DOS}
\dfrac{D_0(\Re)}{2} = \displaystyle \sum_{\vec{p},s\in\Re} \left(-\dfrac{\partial f_0}{\partial E}\right),
\end{equation}
where the factor of 2 appeared since we are summing over all $s$ states. Note that $D_0(U^+)=D_0(D^-)$ and $D_0(U^-)=D_0(D^+)$ due to time-reversal symmetry.

We make the following assumption in Eq. \eqref{semi_sum}
\begin{equation}
\label{cond}
\langle v_x \mu \rangle \approx \langle v_x \rangle \langle \mu \rangle,
\end{equation}
which yields
\begin{equation}
\label{semi_dos}
\begin{aligned}
&\dfrac{D_0(\Re)}{2}\dfrac{\partial \left\langle \mu  \right\rangle }{\partial t} + \dfrac{D_0(\Re)}{2}\langle v_x\rangle\dfrac{\partial \langle {\mu} \rangle}{\partial x} \\&+ \dfrac{qD_0(\Re)}{2}\langle v_x\rangle L_M \dfrac{\partial I_c}{\partial t}
- \dfrac{q}{C_E}\dfrac{D_0(\Re)}{2}\dfrac{\partial Q}{\partial t} \\&= -\displaystyle \sum_{\vec{p},s\in\Re}\,\, \displaystyle \sum_{\vec{p'},s'}S(\vec{p},s\leftrightarrow\vec{p'},s') \left(-\dfrac{\partial f_0}{\partial E}\right){\left(\mu - \mu ' \right)}+\dfrac{i_{ext}}{q}.
\end{aligned}
\end{equation}
Note that the term $i_{ext}/q$ on the right hand side has been added to take into account the total current entering into $\vec{p},s\in\Re$ states of the channel an external contact.

The number of modes within $\vec{p},s\in\Re$ in the channel is given by \cite{Datta_LNE_2012} 
\begin{equation}
\label{NOM}
n_m(\Re)=\dfrac{h D_0|\langle v_x\rangle|}{2L},
\end{equation}
where $|\langle v_x\rangle|$ is the magnitude of $\langle v_x\rangle$ and $L$ is the channel length. 
Note that $|\langle v_x\rangle|$ is same in all four groups for the Rashba Hamiltonian considered here (see Appendix \ref{AppC}). Thus Eq. \eqref{semi_dos} can be written as
\begin{equation}
\label{semi1Dfinal}
\begin{aligned}
&\dfrac{{n_m\left(\Re \right)}}{{\left| {\left\langle {{v_x}} \right\rangle } \right|}}\dfrac{{\partial \left\langle \mu  \right\rangle }}{{\partial t}} + {\mathop{\rm sgn}} \left( {\left\langle {{v_x}} \right\rangle } \right)  n_m\left( { \Re } \right) \dfrac{{\partial \langle \mu \rangle }}{{\partial x}} - \dfrac{q}{{{C_E}}}\frac{{n_m\left( { \Re } \right)}}{{\left| {\left\langle {{v_x}} \right\rangle } \right|}}\frac{{\partial Q}}{{\partial t}}\\&+ {\mathop{\rm sgn}} \left( {\left\langle {{v_x}} \right\rangle } \right) q \, {n_m\left( {\Re } \right)} {L_M}\dfrac{{\partial {I_c}}}{{\partial t}} = \tilde{S}(\Re) + \frac{h}{q}\frac{{{i_{ext}}}}{L},
\end{aligned}
\end{equation}
where
\begin{equation}
\label{scatter1}
\tilde{S}(\Re)=-\frac{h}{L}\sum\limits_{\vec p,s \in \Re } {\mkern 1mu}  {\mkern 1mu} \sum\limits_{\vec p',s'} S (\vec p,s \leftrightarrow \vec p',s')\left( { - \frac{{\partial {f_0}}}{{\partial E}}} \right)\left( {\mu  - \mu '} \right).
\end{equation}
Eq. \eqref{semi1Dfinal} applies to each group in Eq. \eqref{classification} with an average electrochemical potential given by $\left\langle \mu \right\rangle_{\vec{p},s \in U^+} \equiv \mu\left(U^+\right)$, $\left\langle \mu \right\rangle_{\vec{p},s \in D^-} \equiv \mu\left(D^-\right)$, $\left\langle \mu \right\rangle_{\vec{p},s \in U^-} \equiv \mu\left(U^-\right)$, and $\left\langle \mu \right\rangle_{\vec{p},s \in D^+} \equiv \mu\left(D^+\right)$ respectively.

\subsection{Scattering Matrix}
We make the following assumption 
\begin{equation}
\label{cond2}
\langle S \mu \rangle \approx \langle S \rangle \langle \mu \rangle,
\end{equation}
which when applied to the term related to the scattering rate $\tilde{S}(\Re)$ in Eq. \eqref{scatter1} becomes
\begin{equation}
\label{scatter}
\begin{aligned}
\tilde{S}(\Re)= &{\hat S_{\Re  \leftrightarrow {U^ + }}}\left( {\left\langle {\mu \left( {{U^ + }} \right)} \right\rangle  - \left\langle {\mu \left( \Re  \right)} \right\rangle } \right) \\
&+ {\hat S_{\Re  \leftrightarrow {D^ - }}}\left( {\left\langle {\mu \left( {{D^ - }} \right)} \right\rangle  - \left\langle {\mu \left( \Re  \right)} \right\rangle } \right)\\
&+ {\hat S_{\Re  \leftrightarrow {U^ - }}}\left( {\left\langle {\mu \left( {{U^ - }} \right)} \right\rangle  - \left\langle {\mu \left( \Re  \right)} \right\rangle } \right)\\
&+ {\hat S_{\Re  \leftrightarrow {D^ + }}}\left( {\left\langle {\mu \left( {{D^ + }} \right)} \right\rangle  - \left\langle {\mu \left( \Re  \right)} \right\rangle } \right),
\end{aligned}
\end{equation}
with
\begin{equation}
{\hat S_{\Re \leftrightarrow \Re '}}=\frac{h}{L}\sum\limits_{\vec p,s \in \Re } \sum\limits_{\vec p',s'\in\Re '} \left( { - \frac{{\partial {f_0}}}{{\partial E}}} \right) S (\vec p,s \leftrightarrow \vec p',s').
\end{equation}

We can evaluate Eq. \eqref{scatter} for each of the group in Eq. \eqref{classification} i.e. $\Re \equiv U^+$, $D^-$, $U^-$, and $D^+$, which together in the $\{\mu(U^+),\mu(D^-),\mu(U^-),\mu(D^+)\}^T$ basis becomes the following matrix (see Appendix \ref{AppE} for details)
\begin{equation}
\label{scatter_mat}
\begin{aligned}
\left[ S \right] = \left[ {\begin{array}{*{20}{c}}
	{ - {u_1}}&{{{\hat S}_{{U^ + } \leftrightarrow {D^ - }}}}&{{{\hat S}_{{U^ + } \leftrightarrow {U^ - }}}}&{{{\hat S}_{{U^ + } \leftrightarrow {D^ + }}}}\\
	{{{\hat S}_{{U^ + } \leftrightarrow {D^ - }}}}&{ - {u_1 '}}&{{{\hat S}_{{D^ - } \leftrightarrow {U^ - }}}}&{{{\hat S}_{{D^ - } \leftrightarrow {D^ + }}}}\\
	{{{\hat S}_{{U^ + } \leftrightarrow {U^ - }}}}&{{{\hat S}_{{D^ - } \leftrightarrow {U^ - }}}}&{ - {u_2}}&{{{\hat S}_{{U^ - } \leftrightarrow {D^ + }}}}\\
	{{{\hat S}_{{U^ + } \leftrightarrow {D^ + }}}}&{{{\hat S}_{{D^ - } \leftrightarrow {D^ + }}}}&{{{\hat S}_{{U^ - } \leftrightarrow {D^ + }}}}&{ - {u_2 '}}
	\end{array}} \right],
\end{aligned}
\end{equation}

where 

\begin{equation*}
\begin{aligned}
&{u_1} = {{\hat S}_{{U^ + } \leftrightarrow {D^ - }}} + {{\hat S}_{{U^ + } \leftrightarrow {U^ - }}} + {{\hat S}_{{U^ + } \leftrightarrow {D^ + }}},\\
&u_1' = {{\hat S}_{{U^ + }\leftrightarrow {D^ - }}} + {{\hat S}_{{D^ - } \leftrightarrow {U^ - }}} + {{\hat S}_{{D^ - } \leftrightarrow {D^ + }}},\\
&{u_2} = {{\hat S}_{{U^ + } \leftrightarrow {U^ - }}} + {{\hat S}_{{D^ - } \leftrightarrow {U^ - }}} + {{\hat S}_{{U^ - } \leftrightarrow {D^ + }}},\\ 
\text{and }\;&{{u_2'}} = {{\hat S}_{{U^ + } \leftrightarrow {D^ + }}} + {{\hat S}_{{D^ - } \leftrightarrow {D^ + }}} + {{\hat S}_{{U^ - } \leftrightarrow {D^ + }}}.
\end{aligned}
\end{equation*}
The scattering matrix is such that the sum of each column is zero satisfying the charge conservation and the sum of each row is zero satisfying the zero current requirement under equal potential.

In addition, the time-reversal symmetry requires that
\begin{equation}
{{\hat S}_{{U^ + } \leftrightarrow {U^ - }}}={{\hat S}_{{D^ - } \leftrightarrow {D^ + }}} \text{  and  } \hat{S}_{U^+ \leftrightarrow D^+}=\hat{S}_{D^- \leftrightarrow U^-}.
\end{equation}
There are three types of scattering processes considered in the channel: (a) reflection with spin flip $r_{s1} = \hat{S}_{U^+ \leftrightarrow D^-}$ and $r_{s2} = \hat{S}_{U^-\leftrightarrow D^+}$, (b) reflection without spin flip $r=\hat{S}_{U^+ \leftrightarrow U^-}=\hat{S}_{D^- \leftrightarrow D^+}$, and (c) transmission with spin flip $t_s=\hat{S}_{U^+ \leftrightarrow D^+}=\hat{S}_{D^- \leftrightarrow U^-}$. They are given by
\begin{subequations}
\begin{equation}
\label{rs1}
r_{s1} = \frac{h}{L}\sum\limits_{\vec p,s \in U^+ } \sum\limits_{\vec p',s'\in D^-} \left( { - \frac{{\partial {f_0}}}{{\partial E}}} \right) S (\vec p,s \leftrightarrow \vec p',s'),
\end{equation}
\begin{equation}
\label{rs2}
r_{s1} = \frac{h}{L}\sum\limits_{\vec p,s \in U^- } \sum\limits_{\vec p',s'\in D^+} \left( { - \frac{{\partial {f_0}}}{{\partial E}}} \right) S (\vec p,s \leftrightarrow \vec p',s'),
\end{equation}
\begin{equation}
\label{r}
\begin{aligned}
r &= \frac{h}{L}\sum\limits_{\vec p,s \in U^+} \sum\limits_{\vec p',s'\in U^-} \left( { - \frac{{\partial {f_0}}}{{\partial E}}} \right) S (\vec p,s \leftrightarrow \vec p',s'),\\
&= \frac{h}{L}\sum\limits_{\vec p,s \in D^-} \sum\limits_{\vec p',s'\in D^+} \left( { - \frac{{\partial {f_0}}}{{\partial E}}} \right) S (\vec p,s \leftrightarrow \vec p',s'),
\end{aligned}
\end{equation}
and
\begin{equation}
\label{ts}
\begin{aligned}
t_s &= \frac{h}{L}\sum\limits_{\vec p,s \in U^+} \sum\limits_{\vec p',s'\in D^+} \left( { - \frac{{\partial {f_0}}}{{\partial E}}} \right) S (\vec p,s \leftrightarrow \vec p',s'),\\
&= \frac{h}{L}\sum\limits_{\vec p,s \in D^-} \sum\limits_{\vec p',s'\in U^-} \left( { - \frac{{\partial {f_0}}}{{\partial E}}} \right) S (\vec p,s \leftrightarrow \vec p',s').
\end{aligned}
\end{equation}
\end{subequations}

Eq. \eqref{semi1Dfinal} for each group in Eq. \eqref{classification} is given as
\begin{equation}
\label{semi_modelx}
\begin{aligned}
&\frac{1}{\left| {\left\langle {{v_x}} \right\rangle } \right|}\frac{\partial }{{\partial t}}\left\{ {\begin{array}{*{20}{c}}
	{M\tilde \mu \left( {{U^ + }} \right)}\\
	{M\tilde \mu \left( {{D^ - }} \right)}\\
	{N\tilde \mu \left( {{U^ - }} \right)}\\
	{N\tilde \mu \left( {{D^ + }} \right)}
	\end{array}} \right\} + \frac{\partial}{{\partial x}}\left\{ {\begin{array}{*{20}{c}}
	{M\tilde \mu \left( {{U^ + }} \right)}\\
	{ - M\tilde \mu \left( {{D^ - }} \right)}\\
	{ - N\tilde \mu \left( {{U^ - }} \right)}\\
	{N\tilde \mu \left( {{D^ + }} \right)}
	\end{array}} \right\} \\&= \left[ {\begin{array}{*{20}{c}}
	{ - {u_1}}&{{r_{s1}}}&r&{{t_s}}\\
	{{r_{s1}}}&{ - {u_1}}&{{t_s}}&r\\
	r&{{t_s}}&{ - {u_2}}&{{r_{s2}}}\\
	{{t_s}}&r&{{r_{s2}}}&{ - {u_2}}
	\end{array}} \right]\left\{ {\begin{array}{*{20}{c}}
	{\tilde \mu \left( {{U^ + }} \right)}\\
	{\tilde \mu \left( {{D^ - }} \right)}\\
	{\tilde \mu \left( {{U^ - }} \right)}\\
	{\tilde \mu \left( {{D^ + }} \right)}
	\end{array}} \right\}\\& - q L_M { \dfrac{{\partial I_c}}{{\partial t}}} \left\{ {\begin{array}{*{20}{c}}
	{M}\\
	{-M}\\
	{-N}\\
	{N}
	\end{array}} \right\} + \dfrac{q}{\left| {\left\langle {{v_x}} \right\rangle } \right| C_E}\dfrac{{\partial Q}}{{\partial t}}\left\{ {\begin{array}{*{20}{c}}
	M\\	M\\	N\\	N
	\end{array}} \right\} + \frac{h}{q}\left\{ {\begin{array}{*{20}{c}}
	{i_U^ + }\\
	{i_D^ - }\\
	{i_U^ - }\\
	{i_D^ + }
	\end{array}} \right\}.
\end{aligned}
\end{equation}
Note that the electrochemical potentials are referenced with respect to the constant $\mu_0$ i.e. $\tilde{\mu} = \mu - \mu_0$. $i_U^+$, $i_D^-$, $i_U^-$, and $i_D^+$ are the currents per unit length entering into the four groups from an external contact (see Fig. \ref{2}(b)), which are given as \cite{Sayed_SciRep_2016}
\begin{equation}
\label{contact_NM_FM}
\begin{aligned}
&i_U^+ = \frac{q^2}{h} M g_u \left(v_u - \frac{\tilde{\mu}(U^+)}{q}\right),\\ &i_D^+ = \frac{q^2}{h} N g_d \left(v_d - \frac{\tilde{\mu}(D^+)}{q}\right),\\ &i_U^- = \frac{q^2}{h} N g_u \left(v_u -\frac{\tilde{\mu}(U^-)}{q}\right),\\\text{and}\;\;\; &i_D^- = \frac{q^2}{h} M g_d \left(v_d - \frac{\tilde{\mu}(D^-)}{q}\right).
\end{aligned}
\end{equation}
Here, $v_u$ and $v_d$ are up and down spin voltages at the external contact respectively. $g_u$ and $g_d$ are up and down spin conductances per unit mode per unit length of the contact. The contact can be either NM ($g_u=g_d$) or FM ($g_u \neq g_d$). In steady-state, Eq. \eqref{semi_modelx} reduces to our prior model \cite{Sayed_SciRep_2016}.

\subsection{Conversion to Charge-Spin Basis}
The charge and spin voltages and currents in the channel are defined in terms of the four average electrochemical potentials as
\begin{equation}
\label{transform}
\left\{ \begin{array}{l}
I_c{R_B}\\
2{V_s}\\
{I_s}{R_B}\\
2{V_c}
\end{array} \right\}\; = \dfrac{q}{h}R_B\left[ {\begin{array}{*{20}{c}}
	1&-1&-1&1\\
	\alpha &-\alpha &{\alpha }&{ - \alpha }\\
	{\dfrac{1}{\alpha }}&{\dfrac{1}{\alpha }}&{-\dfrac{1}{\alpha }}&{ - \dfrac{1}{\alpha }}\\
	1&{ 1}&{ 1}&1
	\end{array}} \right]\;\left\{ \begin{array}{l}
M\tilde \mu ({U^ + })\\
M\tilde \mu ({D^ - })\\
N\tilde \mu ({U^ - })\\
N\tilde \mu ({D^ + })
\end{array} \right\},
\end{equation}
where $R_B$ is the ballistic resistance of the channel given in Eq. \eqref{bal_res} and $\alpha$ is an angular averaging factor. The derivation of Eq. \eqref{transform} is given in Appendix \ref{AppF}.

We have multiplied the second row of Eq. \eqref{transform} with a factor $0\leq\alpha\leq1$ to take into account the angular distribution of the spin polarization of the eigenstates on the half Fermi circles indicated by $U^+$, $U^-$, $D^+$, and $D^-$ in Fig. \ref{2}(a). The net $z$-spin polarization (or $z$-spin voltage) is expected to be lower by the factor $\alpha$ \cite{Hong_PRB_2012, Hong_SciRep_2016, Sayed_SciRep_2016} which depends on the  distribution of the spin polarization of the eigenstates and the details of the scattering mechanisms. The $\alpha$ factor introduced in Eq. \eqref{transform} appears in Eq. \eqref{Hong_Eqn} to indicate a lowering of the charge current induced spin voltage in the channel from the ideal value due to such angular distribution. Onsager reciprocity requires that the spin current induced charge voltage in the channel will be lowered by the same factor $\alpha$ as shown in Eq. \eqref{Hong_Inv}, which has been taken into account by multiplying $1/\alpha$ to the third row of Eq. \eqref{transform}. In the simplest approximation, the angle between the $z$-axis and the spin polarization of the eigenstates of a particular half Fermi circle in Fig. \ref{2}(a) varies from $-\pi/2$ to $+\pi/2$ which yields $\alpha=2/\pi$. 

Combining Eq. \eqref{semi_modelx} with Eq. \eqref{transform} yields
\begin{equation}
\label{charge_spin_eqn}
\begin{array}{l}
\dfrac{1}{{\left| {\left\langle {{v_x}} \right\rangle } \right|}}\dfrac{\partial}{{\partial t}}\;\left[ {\begin{array}{*{20}{c}}
	0&0&0&1\\
	0&0&\alpha^2&0\\
	0&\dfrac{1}{\alpha^2}&0&0\\
	1&0&0&0
	\end{array}} \right]\left\{ \begin{array}{l}
I_c{R_B}\\
2{V_s}\\
{I_s}{R_B}\\
2{V_c}
\end{array} \right\} + \dfrac{\partial}{{\partial x}}\left\{ \begin{array}{l}
I_c{R_B}\\
2{V_s}\\
{I_s}{R_B}\\
2{V_c}
\end{array} \right\}\; \\ = \;\left[ {\begin{array}{*{20}{c}}
	0&0&0&0\\
	0&0&{ - \dfrac{2 \alpha^2}{{{\lambda _0}}}}&{\dfrac{2\alpha p_0}{{{\lambda _0} }}}\\
	{\dfrac{2}{\alpha\lambda_s '}}&{ - \dfrac{2}{{{\alpha^2\lambda _s}}}}&0&0\\
	{ - \dfrac{2}{\lambda }}&{\dfrac{2}{{\alpha \lambda '}}}&0&0
	\end{array}} \right]\left\{ \begin{array}{l}
I_c{R_B}\\
2{V_s}\\
{I_s}{R_B}\\
2{V_c}
\end{array} \right\}\\ - 2{L_M}\dfrac{{\partial I_c}}{{\partial t}}\left\{ \begin{array}{l}
0\\
0\\
\dfrac{p_0}{\alpha}\\
1
\end{array} \right\} + \dfrac{1}{{\left| {\left\langle {{v_x}} \right\rangle } \right|}}\dfrac{2}{{{C_E}}}\dfrac{{\partial Q}}{{\partial t}}\left\{ \begin{array}{l}
1\\
\alpha {p_0}\\
0\\
0
\end{array} \right\} + \left\{ \begin{array}{l}
{R_B i^c}\\
2\Delta v^s\\
R_B i^s\\
2\Delta v^c
\end{array} \right\},
\end{array}
\end{equation}
with the external contact terms given by
\begin{equation}
\label{cont}
\begin{aligned}
&i^c = i_U^+ + i_D^- + i_U^- + i_D^+,\\
&i^s = \dfrac{1}{\alpha} \left(i_U^+ - i_D^- + i_U^- - i_D^+\right)\\
&\Delta v^c = \frac{R_B}{2}\left( i_U^+ - i_D^- - i_U^- + i_D^+ \right), \text{ and}\\
&\Delta v^s = \frac{\alpha R_B}{2}\left(i_U^+ + i_D^- - i_U^- - i_D^+ \right).
\end{aligned}
\end{equation}
Eq. \eqref{cont} combined with Eq. \eqref{contact_NM_FM} yields Eqs. \eqref{contact_charge_spin} and \eqref{contact_charge_spin1}.

\subsection{Continuity Equation}
The term $\partial Q / \partial t$ on the right hand side of Eq. \eqref{charge_spin_eqn} is related to the charge currents according to the continuity equation given by
\begin{equation}
\label{cont_eqn}
\dfrac{\partial Q}{\partial t} + \dfrac{\partial I_c}{\partial x}=i^c.
\end{equation}

The first row of Eq. \eqref{charge_spin_eqn} combined with Eq. \eqref{cont_eqn} becomes
\begin{equation}
\label{sup_eqn}
\frac{\partial }{{\partial t}}{V_c} = \left( {\frac{{\left| {\left\langle {{v_x}} \right\rangle } \right|{R_B}}}{2} + \frac{1}{{{C_E}}}} \right)\left({i^c}-\frac{\partial }{{\partial x}}{I_c}\right),
\end{equation}
which is the first equation in Eq. \eqref{charge_TL_cont}.

The second row of Eq. \eqref{charge_spin_eqn} combined with Eq. \eqref{cont_eqn} becomes
\begin{equation}
\begin{aligned}
\alpha^2 \frac{{{R_B}}}{{2\left| {\left\langle {{v_x}} \right\rangle } \right|}}\frac{\partial }{{\partial t}}{I_s} &+ \frac{\partial }{{\partial x}}{V_s} =  - \frac{{\alpha^2{R_B}}}{{{\lambda _0}}}{I_s} + \frac{{2\alpha {p_0}}}{{{\lambda _0}}}{V_c} \\&+ \alpha {p_0}\frac{1}{{\left| {\left\langle {{v_x}} \right\rangle } \right|}}\frac{1}{{{C_E}}}\left( {{i^c} - \frac{{\partial {I_c}}}{{\partial x}}} \right) + \Delta {v^s}.
\end{aligned}
\end{equation}
We replace the expression for ${{i^c} - \dfrac{{\partial}}{{\partial x}}}{I_c}$ from Eq. \eqref{sup_eqn} to get the second equation in Eq. \eqref{spin_TL_cont}. For contact conductance $G_0\rightarrow0$, Eqs. \eqref{charge_TL_cont} and \eqref{spin_TL_cont} reduces to Eqs. \eqref{charge_TL} and \eqref{spin_TL} respectively.

\subsection{Mean Free Paths}
We have three distinct mean free paths in Eq. \eqref{charge_spin_eqn}, given by
\begin{equation}
\label{mfps}
\begin{aligned}
&\dfrac{1}{\lambda } = \dfrac{1}{2}\left( {\dfrac{{{r_{s2}}}}{N} + \dfrac{{{r_{s1}}}}{M}} \right) + \dfrac{r}{2}\left( {\dfrac{1}{N}\; + \dfrac{1}{M}} \right),\\
&\dfrac{1}{{{\lambda _0}}} = \dfrac{r + {t_s}}{2} \left( {\dfrac{1}{N} + \dfrac{1}{M}} \right), \;\text{and}\\
&\dfrac{1}{{{\lambda _s}}} = \dfrac{1}{2}\left( {\dfrac{{{r_{s2}}}}{N} + \dfrac{{{r_{s1}}}}{M}} \right) + \dfrac{{t_s}}{2}\left( {\dfrac{1}{N} + \dfrac{1}{M}} \right),\,
\end{aligned}
\end{equation}
where $\lambda$, $\lambda_0$, and $\lambda_s$ determine the series charge resistance $R_c$ (see Eq. \eqref{rc}), the series spin resistance $R_s$ (see Eq. \eqref{rs}), and the shunt spin conductance $G_{sh}$ (see \eqref{gsh}) respectively. Note that $\lambda_s$ depends on the spin-flip processes in the channel and determines the shunt conductance $G_{sh}$ that takes into account the spin relaxation process.

\subsection{Charge-Spin Coupling Coefficients}
The other terms of Eq. \eqref{charge_spin_eqn} are given by
\begin{equation}
\label{spin_charge_coupling}
\begin{array}{l}
\dfrac{1}{{\lambda '}} = \dfrac{1}{2}\left( {\dfrac{{{r_{s2}}}}{N} - \dfrac{{{r_{s1}}}}{M}} \right) + \dfrac{r}{2}\left( {\dfrac{1}{N}\; - \dfrac{1}{M}} \right),\;\text{and}\\
\dfrac{1}{{{{\lambda_s '}}}} = \dfrac{1}{2}\left( {\dfrac{{{r_{s2}}}}{N} - \dfrac{{{r_{s1}}}}{M}} \right) + \dfrac{t_s}{2}\left( {\dfrac{1}{N} - \dfrac{1}{M}} \right),
\end{array}
\end{equation}
representing coupling coefficients between charge and spin. $\lambda_s '$ and $\lambda '$ cause a charge induced spin signal and spin induced charge signal respectively. Note that the first terms of Eq. \eqref{spin_charge_coupling} indicate a purely scattering induced spin-charge coupling even if $M=N$ (i.e. $p_0=0$) since $r_{s1}$ and $r_{s2}$ are two independent parameters and there could be situations where $r_{s1}\neq r_{s2}$.

In this paper, we restrict ourselves to SML caused by difference between $M$ and $N$ (i.e. $p_0\neq0$). We can eliminate the first terms in Eq. \eqref{spin_charge_coupling} by assuming either of the followings:
\begin{subequations}
\label{assump12}
\begin{equation}
\label{assump1}
r_{s1}=r_{s2}=r_s,
\end{equation}
\begin{equation}
\label{assump2}
\text{or,}\;\;\dfrac{r_{s1}}{M}=\dfrac{r_{s2}}{N}.
\end{equation}
\end{subequations}

The first assumption Eq. \eqref{assump1} when applied in Eq. \eqref{spin_charge_coupling} yields
\begin{equation}
\label{spin_charge_coupling1}
\begin{array}{l}
\dfrac{1}{{\lambda '}} = (r_s+r) \left( {\dfrac{{{1}}}{N} - \dfrac{{{1}}}{M}} \right)=\dfrac{p_0}{\lambda},\;\text{and}\\
\dfrac{1}{{{{\lambda_s '}}}} = (r_s + t_s) \left( {\dfrac{{{1}}}{N} - \dfrac{{{1}}}{M}} \right)=\dfrac{p_0}{\lambda_s},
\end{array}
\end{equation}
which in turn gives $\lambda_r=\lambda$ and $\lambda_t=\lambda_s$ in Eqs. \eqref{etc} and \eqref{ets} respectively.

The second assumption Eq. \eqref{assump2} when applied in Eq. \eqref{spin_charge_coupling} yields
\begin{equation}
\begin{aligned}
\label{spin_charge_coupling2}
&\dfrac{1}{\lambda '} = r\left( {\dfrac{1}{N}\; - \dfrac{1}{M}} \right) = \dfrac{p_0}{\lambda _r},\;\text{and}\\
&\dfrac{1}{\lambda_s '} = t_s\left( {\dfrac{1}{N} - \dfrac{1}{M}} \right) = \dfrac{p_0}{\lambda _t}.
\end{aligned}
\end{equation}

\subsection{Comments on the Assumptions}
The assumptions in Eqs. \eqref{cond}, \eqref{cond2}, and \eqref{assump12} result in an effective change in the transmission line model parameters in Eqs. \eqref{params},  but does not change the models in Eqs. \eqref{charge_TL}, \eqref{spin_TL} (Fig. \ref{1}) and Eqs. \eqref{charge_TL_cont}, \eqref{spin_TL_cont} (Fig. \ref{3}) themselves. The assumptions made to derive the model can be revisited as the field evolves. However, several predictions from our model for steady-state \cite{Sayed_SciRep_2016} have already received support from experiments \cite{Pham_APL_2016,Pham_NanoLett_2016,Koo_New} suggesting that the assumptions are within the reasonable limits.

\section{Summary}
We have proposed a two component (charge and $z$-component of spin) transmission line model for channels with spin-momentum locking (SML) which is a new addition to our SPICE compatible multi-physics model library \cite{ModApp, Camsari_SciRep_2015, Sayed_SciRep2_2016}. The model enables easy analysis of complex geometries involving materials with spin-orbit coupling (SOC) observed in diverse classes of materials e.g. topological insulators, heavy metals, oxide interfaces, and narrow bandgap semiconductors. The model is derived from a four-component diffusion equation obtained from the Boltzmann transport equation assuming linear response and elastic scattering in the channel. The four-component diffusion equation uses four average electrochemical potentials based on a classification depending on the sign of $z$-component of spin (up ($U$) or down ($D$)) and the sign of $x$-component of group velocity ($+$ or $-$).  Such classification can be viewed as an extension of the Valet-Fert equation \cite{Valet_Fert_1993} which uses two electrochemical potentials for $U$ and $D$ states. For a normal metal channel, the time-dependent model presented here decouples into (i) the well-known transmission line model for charge transport in quantum wires \cite{Burke_TNANO_2002, Burke_TNANO_2003, Sayeef_IEEE_2005} and (ii) a time-dependent version of Valet-Fert equation \cite{Valet_Fert_1993} for spin transport. We first derive several results on charge-spin interconversion starting from our model in steady-state. The steady-state results show good agreement with existing experiments on diverse materials. We then study the phenomenon of spin-charge separation using our full time-dependent model, especially in the materials with SOC exhibiting SML. Our model shows the expected spin-charge separation with two distinct velocities for charge and spin, which persist even in channels exhibiting SML. However, we show that the lower velocity signal is purely spin while the higher velocity signal is largely charge with an additional spin component proportional to the degree of SML.

\begin{acknowledgments}
	This work was in part supported by FAME, one of six centers of STARnet, a Semiconductor Research Corporation (SRC) program sponsored by MARCO and DARPA and  in part by ASCENT, one of six centers in JUMP, a SRC program sponsored by DARPA.
\end{acknowledgments}

\appendix
\section{Steady-State Results}
\label{App_Rmat}
\noindent\textit{This appendix provides the details of the derivation of Eq. \eqref{R_Mat} and Eq. \eqref{IREE_length}.}\\

\subsection{Derivation of the Resistance Matrix}

\textit{Potentiometric NM Contacts}: Eqs. \eqref{contact_charge_spin} and \eqref{contact_charge_spin1} for $p_f=0$ becomes
\begin{subequations}
	\begin{alignat}{2}
	\label{ic_NM}
	&{i^c} = {G_0}\left( {{v_c} - {V_c}} \right),\\
	\label{is_NM}
	&{i^s} = \frac{{{G_0}}}{{{\alpha ^2}}}\left( {{v_s} - {V_s}} \right),\\
	\label{dvc_NM}
	&\Delta {v^c} = \frac{{{G_0}{R_B}}}{{2\alpha }}\left( {{p_0}{v_s} - \frac{{\alpha I_c{R_B}}}{2}} \right),\\
	\label{dvs_NM}
	\text{and}\;\;&\Delta {v^s} = \frac{{\alpha {G_0}{R_B}}}{2}\left( {{p_0}{v_c} - \frac{{\alpha {I_s}{R_B}}}{2}} \right).
	\end{alignat}
\end{subequations}

We apply Eqs. \eqref{ic_NM} and \eqref{is_NM} in Eqs. \eqref{dvc_NM} and \eqref{dvs_NM} respectively, which yields
\begin{equation}
\label{dvv}
\begin{array}{l}
\Delta {v^c} = \dfrac{{\alpha {p_0}}}{{2{G_B}}}{i^s} + {p_0}\dfrac{{{G_0}{R_B}}}{{2\alpha }}{V_s} - {R_{cont}^c}{I_c},\\
\Delta {v^s} = \dfrac{{\alpha {p_0}}}{{2{G_B}}}{i^c} + {p_0}\dfrac{{\alpha {G_0}{R_B}}}{2}{V_c} - {R_{cont}^s}{I_s}.
\end{array}
\end{equation}

We assume that the contact conductance per unit length $G_0$ is very low such that the potentiometric condition in Eq. \eqref{pot_cond} is satisfied. Thus, we have 
\begin{equation*}
\begin{aligned}
&R_c \gg R_{cont}^c,\;\;\; R_s \gg R_{cont}^s,\\
&\eta_c \gg \frac{G_0 R_B}{2\alpha}, \;\;\text{and}\;\; \eta_s \gg \frac{\alpha G_0 R_B}{2}.
\end{aligned}
\end{equation*}
Under this condition, we combine Eq. \eqref{steady_state_TL} with Eq. \eqref{dvv} for $i^c=0$, which yields
\begin{subequations}
	\begin{alignat}{4}
	\label{stead_Ic}
	&\dfrac{d}{{d x}}I_c = 0,\\
	\label{stead_Vc}
	&\dfrac{d}{d x}{V_c} = - R_c\,I_c + {p_0}{\eta_c}{V_s} + \dfrac{{\alpha {p_0}}}{{2{G_B}}}{i^s},\\
	\label{stead_Is}
	&\dfrac{d}{{d x}}{I_s} =  -G_{sh}{V_s} + p_0 \gamma_s I_c + i^s,\\
	\label{stead_Vs}
	\text{and}\;\;&\dfrac{d}{{d x}}{V_s} =  - {R_s}{I_s} + p_0 \eta_s {V_c}.
	\end{alignat}
\end{subequations}

\textit{Uniform Spin Voltage}: We assume that the spin voltage $V_s$ is uniform in the channel region of interest (from $x=0$ to $x=L$ where $L$ is the channel length)
\begin{equation}
\dfrac{d}{dx}V_s = 0.
\end{equation}
Thus from Eq. \eqref{stead_Vs} we can write
\begin{equation}
\label{unif_spin}
{I_s} = \frac{{2{p_0}{G_B}}}{\alpha }V_c,
\end{equation}
where we have used the definitions in Eqs. \eqref{rs} and \eqref{ets}.

Differentiating both sides of Eq. \eqref{unif_spin} with respect to $x$ and combining with Eqs. \eqref{stead_Vc} and \eqref{stead_Is} yields
\begin{equation*}
- {G_{sh}}{V_s} + {p_0}{\gamma _s}{I_c} + {i^s} = \frac{{2{p_0}{G_B}}}{\alpha }\left( { - {R_c}{\mkern 1mu} {I_c} + {p_0}{\eta _c}{V_s} + \frac{{\alpha {p_0}}}{{2{G_B}}}{i^s}} \right),
\end{equation*}
which in conjunction with the definitions in Eq. \eqref{params} yields
\begin{equation}
\label{first_row}
{V_s} = \dfrac{{\alpha {p_0}}}{{2{G_B}}}\dfrac{{\left( {\dfrac{1}{{{\lambda _t}}} + \dfrac{1}{\lambda }} \right)}}{{\left( {\dfrac{1}{{{\lambda _s}}} + \dfrac{{p_0^2}}{{{\lambda _r}}}} \right)}}{I_c} + \dfrac{{{\alpha ^2}\left( {1 - p_0^2} \right)}}{{4{G_B}}}\dfrac{1}{{\left( {\dfrac{1}{{{\lambda _s}}} + \dfrac{{p_0^2}}{{{\lambda _r}}}} \right)}}{i^s}.
\end{equation}

From Eq. \eqref{is_NM} we have
\begin{equation}
{V_s} = {v_s} - \frac{{{\alpha ^2}{i^s}}}{{{G_0}}},
\end{equation}
which when combined with Eq. \eqref{first_row}, becomes
\begin{equation}
\label{first_row1}
\begin{aligned}
{v_s} &= \dfrac{{\alpha {p_0}}}{{2{G_B}}}\dfrac{{\left( {\dfrac{1}{{{\lambda _t}}} + \dfrac{1}{\lambda }} \right)}}{{\left( {\dfrac{1}{{{\lambda _s}}} + \dfrac{{p_0^2}}{{{\lambda _r}}}} \right)}}{I_c} \\
&+ \left(\dfrac{{{\alpha ^2}\left( {1 - p_0^2} \right)}}{{4{G_B}}}\dfrac{1}{{\left( {\dfrac{1}{{{\lambda _s}}} + \dfrac{{p_0^2}}{{{\lambda _r}}}} \right)}}+\dfrac{\alpha^2}{G_0}\right){i^s}.
\end{aligned}
\end{equation}

\textit{Charge Voltage in the Channel}: Let us assume that the contact 1 is located at $x=0$ and contact 2 is located at $x=L$ as shown in Fig. \ref{4}, where $L$ is the channel length between the two contacts. The charge voltage in the channel at $x=0$ and $x=L$ are $V_{c1}$ and $V_{c2}$ respectively. According to \eqref{stead_Ic}, $I_c$ is constant along the channel. Since $V_s$ and $i^s$ is also uniform along the channel, we can integrate Eq. \eqref{stead_Vc} from $x=0$ to $x=L$ as
\begin{equation}
\label{Vdiff}
V_{c1} - V_{c2} = \dfrac{L}{\lambda }{R_B}{I_c} - \frac{{2{p_0L}}}{{\alpha {\lambda _r}}}{V_s} - \frac{{\alpha {p_0}}}{{2{G_B}}}{i^s}L.
\end{equation}

From Eq. \eqref{ic_NM}, we can write equation for $V_{c1}$ in terms of the terminal voltage $v_{c1}$ of contact 1 at $x=0$ as
\begin{equation}
\label{vc1}
V_{c1} = v_{c1} - \frac{{{I_c}}}{{{G_0''}}}.
\end{equation}
noting that $i^c=I_c$. Here, $G_0''$ is the contact conductance of contact 1. Similarly, we can write an equation for contact 2 noting that $i^c=-I_c$ given by
\begin{equation}
\label{vc2}
V_{c2} = v_{c2} + \frac{{{I_c}}}{{{G_0''}}},
\end{equation}
where we assume that contact 1 and 2 has same conductance.

We combine Eq. \eqref{Vdiff} with Eqs. \eqref{first_row}, \eqref{vc1}, and \eqref{vc2} to have
\begin{equation}
\label{second_row}
\begin{aligned}
{v_{c1}} - {v_{c2}} =\left[ {\left\{ {\frac{L}{\lambda } - \frac{{p_0^2L}}{{{\lambda _r}}}\frac{{\left( {\frac{1}{{{\lambda _t}}} + \frac{1}{\lambda }} \right)}}{{\left( {\frac{1}{{{\lambda _s}}} + \frac{{p_0^2}}{{{\lambda _r}}}} \right)}}} \right\}{R_B} + \frac{2}{{{G_0}^{\prime \prime }}}} \right]{I_c}\\- \frac{{\alpha {p_0}}}{{2{G_B}}}\frac{{\left( {\dfrac{1}{{{\lambda _r}}} + \dfrac{1}{{{\lambda _s}}}} \right)}}{{\left( {\dfrac{1}{{{\lambda _s}}} + \dfrac{{p_0^2}}{{{\lambda _r}}}} \right)}}{i^s}L.
\end{aligned}
\end{equation}

\textit{Scattering Condition}: We assume that the reflection with spin-flip scattering mechanism is dominant in the channel
\begin{equation}
r_{s1,2}\gg r,\;t_s.
\end{equation}

This condition in Eqs. \eqref{mfps}, \eqref{spin_charge_coupling1}, and \eqref{spin_charge_coupling} yields
\begin{equation}
\label{scat_cond}
\dfrac{1}{\lambda}\approx \dfrac{1}{\lambda_s},\;\;\text{and}\;\;\dfrac{1}{\lambda},\dfrac{1}{\lambda_s}\gg \dfrac{1}{\lambda_0},\dfrac{1}{\lambda_r},\dfrac{1}{\lambda_t}.
\end{equation}

Applying Eq. \eqref{scat_cond} in Eqs. \eqref{second_row} and \eqref{first_row1} we have the first and second rows of Eq. \eqref{R_Mat} respectively by setting $\Delta V_c=v_{c1}-v_{c2}$, $G_0'=G_0L$ and $i^s_{tot}=i^sL$.

\subsection{Derivation of the IREE Length}
We start from the first row of Eq. \eqref{R_Mat} given by
\begin{equation*}
\Delta V_c= \left({\frac{L}{{{G_B}\lambda }}}+\dfrac{2}{G_0''}\right)I_c - \frac{{\alpha {p_0}}}{{2{G_B}}}i^s_{tot}.
\end{equation*}
We assume that the channel resistance is much larger than the contact resistances, i.e.  $L/(G_B\lambda) \gg 2/G_0''$. This yields
\begin{equation}
\Delta V_c= {\frac{L}{{{G_B}\lambda }}}I_c - \frac{{\alpha {p_0}}}{{2{G_B}}}i^s_{tot}.
\end{equation}

We apply the short circuit condition $v_{c1}=v_{c2}$ at contacts 1 and 2 i.e. $\Delta V_c = 0$. Thus we get
\begin{equation}
I_c = \dfrac{\alpha p_0 \lambda}{2}\dfrac{i^s_{tot}}{L}.
\end{equation}

The IREE length is given by Eq. \eqref{defIREE} as
\begin{equation}
\lambda_{IREE}=\dfrac{J_c}{J_s}=\dfrac{I_c/w}{I_s/(wL)}=\dfrac{\alpha p_0 \lambda}{2},
\end{equation}
which is the expression in Eq. \eqref{IREE_length1}. Applying $\alpha = 2/\pi$ yields Eq. \eqref{IREE_length}.

\section{Simulation Setup}
\label{App_Hong}

\noindent\textit{This appendix provides the details of the simulation setup in SPICE that was used to analyze steady-state results of charge-spin interconversion in Section III.}\\

\begin{figure}
	\centering
	\includegraphics[width=0.49 \textwidth]{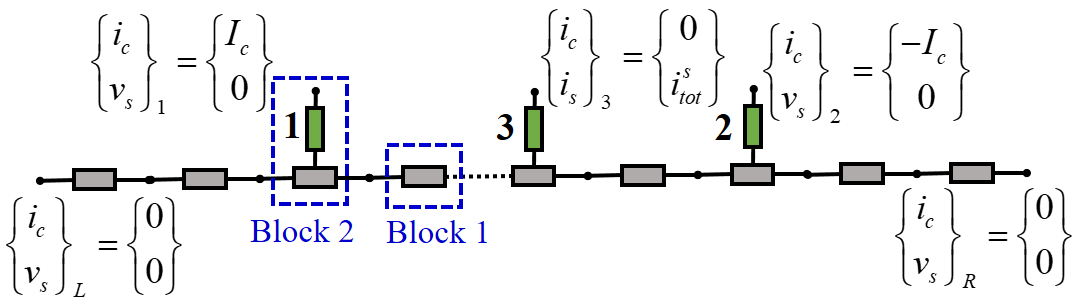}
	\caption{SPICE setup for dc simulation. The setup connects the transmission line circuit models in a distributed manner. Block 1 corresponds to model in Fig. \ref{1} and Block 2 corresponds to model in Fig. \ref{3}. All the external contact have $p_f=0$. All terminals are two component: charge ($c$) and spin ($s$). All indicated spin terminals are grounded except the spin terminal of contact 2.}\label{3_7}
\end{figure}

We have discretized the structure in Figs. \ref{4}(a), \ref{4}(c), and \ref{5}(a) into 100 small sections and represented each of the small sections with the corresponding circuit model. For example, Block 1 and block 2 indicated in Fig. \ref{3_7} are represented with the models in Figs. \ref{1} and \ref{3} respectively. Note that each of the nodes in Fig. \ref{3_7} are two component: charge ($c$) and $z$-component of spin ($s$). We have connected the charge and spin terminals of the models for all the small sections in a modular fashion using standard circuit rules as shown in Fig. \ref{3_7}. We perform a dc simulation in SPICE. Note that during dc simulation in SPICE, capacitors and inductors automatically become open and short circuit respectively and correspond to the stead-state ($\partial/\partial t \rightarrow 0$) form in Eq. \eqref{steady_state_TL}.

The contacts (1, 2, and 3) in this discussion are point contacts. The contact polarizations $p_f=0$ and conductances are in the potentiometric limit with $G_0\approx0.05G_B$. We set the total number of modes $M+N$ in the channel to be 100. We have assumed that the reflection with spin-flip scattering mechanism is dominant in the channel i.e. $r_{s1,2}\gg r,t_s$. The scattering rate per unit mode was set to 0.04 per lattice point.

We apply the charge open and spin ground boundary condition at the two boundaries given by
\begin{equation}
{\left\{ {\begin{array}{*{20}{c}}
		{{i_c}}\\
		{{v_s}}
		\end{array}} \right\}_L} = \left\{ {\begin{array}{*{20}{c}}
	0\\
	0
	\end{array}} \right\},\,\,\,\,\text{and}\,\,{\left\{ {\begin{array}{*{20}{c}}
		{{i_c}}\\
		{{v_s}}
		\end{array}} \right\}_R} = \left\{ {\begin{array}{*{20}{c}}
	0\\
	0
	\end{array}} \right\}.
\end{equation}
Here, $i_c$ and $v_s$ indicates boundary charge current and boundary spin voltage. Indices $L$ and $R$ indicate left and right boundaries respectively.

\textit{Setup in Fig. \ref{4}(a)}: For setup in Fig. \ref{4}(a), we apply a current $I_c$ at the charge terminals of contacts 1 and 2, given by
\begin{equation}
{\left\{ {\begin{array}{*{20}{c}}
		{{i_c}}\\
		{{v_s}}
		\end{array}} \right\}_1} = \left\{ {\begin{array}{*{20}{c}}
	{{I_c}}\\
	0
	\end{array}} \right\}\;\;\text{and}\;\;{\left\{ {\begin{array}{*{20}{c}}
		{{i_c}}\\
		{{v_s}}
		\end{array}} \right\}_2} = \left\{ {\begin{array}{*{20}{c}}
	{ - {I_c}}\\
	0
	\end{array}} \right\}.
\end{equation}
where indices $1$ and $2$ represent contacts 1 and 2 respectively. The spin terminals of contacts 1 and 2 are grounded to take into account the spin relaxation within the contact. Both charge and spin terminals of contact 3 are open and we observe open circuit spin voltage at the spin terminal. The boundary condition at contact 3:
\begin{equation}
{\left\{ {\begin{array}{*{20}{c}}
		{{i_c}}\\
		{{i_s}}
		\end{array}} \right\}_3} = \left\{ {\begin{array}{*{20}{c}}
	0\\
	0
	\end{array}} \right\},
\end{equation}
where index $3$ indicate the contact 3.

\textit{Setup in Fig. \ref{4}(c)}: For the setup in Fig. \ref{4}(c), the charge terminal of contact 3 is open and we apply a current $i^s_{tot}$ at the spin terminal, given by
\begin{equation}
{\left\{ {\begin{array}{*{20}{c}}
		{{i_c}}\\
		{{i_s}}
		\end{array}} \right\}_3} = \left\{ {\begin{array}{*{20}{c}}
	0\\
	{i_{tot}^s}
	\end{array}} \right\}.
\end{equation}
The spin terminals of contacts 1 and 2 are grounded to take into account the spin relaxation within the contact and charge terminals are kept open. Here we observe the open circuit voltage difference between the charge terminals of contacts 1 and 2. The boundary conditions are:
\begin{equation}
{\left\{ {\begin{array}{*{20}{c}}
		{{i_c}}\\
		{{v_s}}
		\end{array}} \right\}_1} = \left\{ {\begin{array}{*{20}{c}}
	0\\
	0
	\end{array}} \right\}\;\;\text{and}\;\;{\left\{ {\begin{array}{*{20}{c}}
		{{i_c}}\\
		{{v_s}}
		\end{array}} \right\}_2} = \left\{ {\begin{array}{*{20}{c}}
	0\\
	0
	\end{array}} \right\}.
\end{equation}

\textit{Setup in Fig. \ref{5}(a)}: For the setup in Fig. \ref{5}(a), charge terminal of contact 3 is open and we inject a current $i^s_{tot}$ through the spin terminal. The boundary condition at contact 3:
\begin{equation}
{\left\{ {\begin{array}{*{20}{c}}
		{{i_c}}\\
		{{i_s}}
		\end{array}} \right\}_3} = \left\{ {\begin{array}{*{20}{c}}
	0\\
	{i_{tot}^s}
	\end{array}} \right\}.
\end{equation}
We short circuit the charge terminals of contacts 1 and 2 and observe the short circuit charge current $I_c$ flowing in the channel induced by $i^s_{tot}$. The spin terminals of contacts 1 and 2 are grounded to take into account the spin relaxation within the contact. The boundary conditions at contacts 1 and 2 are:
\begin{equation}
{\left\{ {\begin{array}{*{20}{c}}
		{{v_c}}\\
		{{v_s}}
		\end{array}} \right\}_1} = {\left\{ {\begin{array}{*{20}{c}}
		{{v_c}}\\
		0
		\end{array}} \right\}_2} \;\;\text{and}\;\;{\left\{ {\begin{array}{*{20}{c}}
		{{v_c}}\\
		{{v_s}}
		\end{array}} \right\}_2} = {\left\{ {\begin{array}{*{20}{c}}
		{{v_c}}\\
		0
		\end{array}} \right\}_1}.
\end{equation}

\section{Parameter Estimations}
\label{App_IREE}
\noindent \textit{This appendix provides the details of the estimations made for the IREE lengths on diverse materials using Eq. \eqref{IREE_length}.}

\subsection{Estimations}

\subsubsection{Estimation of Fermi Wave Vector ($k_F$)}
We estimate the Fermi wave vector $k_F$ of the channel from the electron density $n_{3D}$ (units of m$^{-3}$) or $n_{2D}$ (units of m$^{-2}$), using the following expressions
	\begin{subequations}
		\begin{equation}
		\label{3DkF}
		{{k}_{F}}=\sqrt[3]{3{{\pi }^{2}}{{n}_{3D}}},
		\end{equation}
		\begin{equation}
		\label{2DkF}
		{{k}_{F}}=\sqrt{2 \pi n_{2D}}.
		\end{equation}
	\end{subequations}
Estimations are summarized in Table \ref{table_kF}.
	
	\begin{table}
		\begin{center}
			\caption{Estimation of $k_F$.}
			\label{table_kF}
			\begin{tabular}{||c | c | c | c | c ||} 
				\hline
				Material & Electron Density & $k_F$ [nm$^{-1}$] \\
				\hline\hline
				Ag$|$Bi$^\dag$ & $5.86\times {{10}^{28}}\,{{\text{m}}^{-3}}$ \cite{Ashcroft1976}  & 12 (Eq. \eqref{3DkF})\\
				\hline
				Cu$|$Bi$^\dag$ & $8.49\times {{10}^{28}}\,{{\text{m}}^{-3}}$ \cite{Ashcroft1976} & 13.6 (Eq. \eqref{3DkF})\\
				\hline
				LAO$|$STO & $2.6\times {{10}^{17}}\,{{\text{m}}^{-2}}$ \cite{FertLAOSTO2016} & 1.278 (Eq. \eqref{2DkF})\\
				\hline
				Bi$_2$Se$_3$ & $4\times {{10}^{17}}\,{{\text{m}}^{-2}}$ \cite{SmarthPRL2016} & 1.59 (Eq. \eqref{2DkF})\\
				\hline
			\end{tabular}
		\end{center}
		\begin{flushleft}
			$^\dag${\footnotesize We have used the electron densities of Ag and Cu respectively as they are the most conductive layer in the corresponding bi-layer.}
		\end{flushleft}
	\end{table}
\subsubsection{Estimation of Ballistic Conductance ($G_B$)}
We estimate the total number of modes in the 3D channel using the following expression
	\begin{equation}
	\label{totM}
	M+N=\frac{{k_F^2wt}}{{2\pi}},
	\end{equation}
	where $w$ is the width and $t$ is the thickness. For a 2D channel the expression is
	\begin{equation}
	\label{totM2D}
	M+N=\frac{2k_F w}{\pi}.
	\end{equation}
	
We estimate the ballistic conductance of the channel using the following expression
	\begin{equation}
	\label{balcond}
	G_B=\dfrac{q^2}{h}(M+N).
	\end{equation}
	Estimations are summarized in Table \ref{table_GB}.

\begin{table}
	\begin{center}
		\caption{Estimation of $G_B$.}
		\label{table_GB}
		\begin{tabular}{||c | c | c | c | c ||} 
			\hline
			Material & $w$ [$\mu$m] & $t$ [nm] &$M+N$ & $G_B$\\&&&&(Eq. \eqref{balcond})\\
			\hline\hline
			Ag$|$Bi$^\dag$ & 400 \cite{Fert_NatComm_2016} & 5 \cite{Fert_NatComm_2016} & $4.6\times10^7$  & 1.77 kS\\&&&(Eq. \eqref{totM})&\\
			\hline
			Cu$|$Bi$^\dag$ & 0.15 \cite{IssaCuBi2016} & 20 \cite{IssaCuBi2016} & $8.8\times10^4$ & 3.4 S\\&&&(Eq. \eqref{totM})&\\
			\hline
			LAO$|$STO & 400 \cite{FertLAOSTO2016} & - & $3.3\times10^5$ & 12.6 S\\&&&(Eq. \eqref{totM2D})&\\
			\hline
			Bi$_2$Se$_3$ & 1000 \cite{SmarthPRL2016} & 9$^{\dag\dag}$ \cite{SmarthPRL2016} & $3.6\times10^6$ & 140 S\\&&&(Eq. \eqref{totM})&\\
			\hline
		\end{tabular}
	\end{center}
	\begin{flushleft}
		$^\dag${\footnotesize We have used the thicknesses of the most conductive layer (Ag and Cu respectively) to estimate $G_B$ considering bulk conduction.}\\
		$^{\dag\dag}${\footnotesize We considered 6 quintuple layer (QL) sample in Ref. \cite{SmarthPRL2016}. 1 QL $\approx$ 1.5 nm.}
	\end{flushleft}
\end{table}

\subsubsection{Estimation of Mean Free Path ($\lambda$)}
The mean free path ($\lambda$) is estimated from the measured sheet resistance $R_S$ of the sample using the following expression:
	\begin{equation}
	\label{mfp}
	\frac{{{R_S}}}{w} = \frac{1}{{{G_B}\lambda }},
	\end{equation}
	or from the resistivity $\rho$ of the sample using the following expression:
	\begin{equation}
	\label{mfp_rho}
	\frac{\rho }{{wt}}=\frac{1}{{{G_B}\lambda }}.
	\end{equation}
	Estimations are summarized in Table \ref{table_lamda}.

\begin{table}
	\begin{center}
		\caption{Estimation of $\lambda$.}
		\label{table_lamda}
		\begin{tabular}{||c | c | c | c | c ||} 
			\hline
			Material & $R_S$ [$\Omega/\square$] & $\rho$ [$\mu\Omega$-cm] &$\lambda$ [nm]\\
			\hline\hline
			Ag$|$Bi & $10$$^\dag$ & - & 22.6 (Eq. \eqref{mfp}) \\
			\hline
			Cu$|$Bi & - & $100$$^{\ddagger}$ & 0.88 (Eq. \eqref{mfp_rho})\\
			\hline
			LAO$|$STO & $176$$^{\dag\dag}$ & - & 180.8 (Eq. \eqref{mfp})\\
			\hline
			Bi$_2$Se$_3$ & - & 2000$^{\ddagger\ddagger}$ & 3.2 (Eq. \eqref{mfp_rho})\\
			\hline
		\end{tabular}
	\end{center}
	\begin{flushleft}
		$^\dag${\footnotesize Taken from Fig. 3(a) of Ref. \cite{Fert_NatComm_2016} for Ag$|$Bi sample with 5 nm Ag.}\\
		$^{\dag\dag}${\footnotesize Taken from Fig. 1(d) of Ref. \cite{FertLAOSTO2016} for LAO$|$STO at 7K.}\\
		$^{\ddagger}${\footnotesize $\rho$ of Bi layer was used which was taken from Ref. \cite{IssaCuBi2016}.}\\
		$^{\ddagger\ddagger}${\footnotesize Taken from Fig. 2(b) of Ref. \cite{SmarthPRL2016} at $\sim$300K.}
	\end{flushleft}
\end{table}

\subsubsection{Estimation of Degree of Spin-Momentum Locking ($p_0$)}
For Ag$|$Bi, Cu$|$Bi, and LAO$|$STO, the degree of SML $p_0$ is estimated using the Rashba coupling coefficient ($v_0=\alpha_R/\hbar$) and the Fermi velocity ($v_F$) of the materials, using Eq. \eqref{RashbaSML}. For Bi$_2$Se$_3$, $p_0$ is estimated using spin current density reported from spin pumping ($J_s$), measured inverse effect voltage $\Delta V_c$, and Eq. \eqref{Hong_Inv}. The estimations are summarized in Table \ref{table_p0}.

\begin{table}
	\begin{center}
		\caption{Estimation of $p_0$.}
		\label{table_p0}
		\begin{tabular}{||c | c | c | c | c ||} 
			\hline
			Material & $\alpha_R$ [eV$\cdot\AA$] & $v_F$ [$\times10^6$ m$\cdot$s$^{-1}$] &$p_0$\\
			\hline\hline
			Ag$|$Bi & 0.56 \cite{Fert_NatComm_2016} & 1.39 (Ag) \cite{Ashcroft1976} & 0.05 \\&&1.87 (Bi) \cite{Ashcroft1976}&\\
			\hline
			Cu$|$Bi & 0.56 \cite{IssaCuBi2016} & 1.57 (Cu) \cite{Ashcroft1976} & 0.05\\&&1.87 (Bi) \cite{Ashcroft1976}&\\
			\hline
			LAO$|$STO & 0.03 \cite{FertLAOSTO2016} & 0.074$^\dag$ & 0.0616\\
			\hline
			\hline
			\hline
			\hline
			Material & $\Delta V_c$ [$\mu$V] & $J_s$ [A-m$^{-2}$] &$p_0$\\
			\hline
			Bi$_2$Se$_3$ & 40$^{\ddagger}$ \cite{SmarthPRL2016} & 1.39$\times$10$^5$ $^{\ddagger\ddagger}$ & 0.025\\&&&(Eq. \eqref{Hong_Inv}).\\
			\hline
		\end{tabular}
	\end{center}
	\begin{flushleft}
		$^\dag${\footnotesize We estimate the Fermi velocity using ${{v}_{F}}=\frac{\hbar {{k}_{F}}}{{{m}^{*}}}$. $m^*\approx$ 2 $\times$ $9.1\times {{10}^{-31}}$ kg as reported in Ref. \cite{FertLAOSTO2016}.}\\
		$\ddagger${\footnotesize Inverse effect induced voltage due to spin-pumping at 3 GHz.}\\
		$^{\ddagger\ddagger}${\footnotesize Taken from table 1 in supplementary information of Ref. \cite{SmarthPRL2016} for 6 QL. The sample dimension is 1 mm $\times$ 5 mm \cite{SmarthPRL2016}, which yields $i^s_{tot}\approx0.695$ A.}
	\end{flushleft}
\end{table}

\subsection{Derivation of Eq. \eqref{RashbaSML}}

We start from the eigenstates in Eq. \eqref{E_eig} of the Rashba Hamiltonian in Eq. \eqref{Rashba}, given by
\[E=\dfrac{{p'}^2}{2m}- s\, v_0 {p'}+U_E.\]
with ${p'}=|\vec{p}-q\vec{A}|$. Solutions for ${p'}$ are given by
\begin{equation*}
\begin{array}{l}
{{p'}_1}(s) = s\,m{v_0} + \sqrt {{m^2}v_0^2 + 2m(E-U_E)}, \\
{{p'}_2}(s) = s\,m{v_0} - \sqrt {{m^2}v_0^2 + 2m(E-U_E)},
\end{array}
\end{equation*}
noting that $s^2=1$. Here, ${{p'}_1}(s=+1)$ and ${{p'}_1}(s=-1)$ correspond to $M$ and $N$ respectively. Similarly ${{p'}_2}(s=-1)$ and ${{p'}_2}(s=+1)$ correspond to $M$ and $N$ respectively. Thus the degree of SML $p_0$ is given by
\begin{equation}
\label{deg_Rashba_SML}
\begin{aligned}
p_0(E_F) &= \dfrac{{{p'}_1}(s=+1)-{{p'}_1}(s=-1)}{{{p'}_1}(s=+1)+{{p'}_1}(s=-1)} \\&= \dfrac{{{v_0}}}{{\sqrt {v_0^2 + \dfrac{{2(E_F-U_E)}}{m}} }}.
\end{aligned}
\end{equation}

Assuming $U_E=0$ and applying $E_F=\dfrac{1}{2}mv_F^2$ we have the expression in Eq. \eqref{RashbaSML}.

\section{Charge and Spin Velocities}
\label{AppA}
\noindent
\textit{This appendix provides the derivation of eigenvalues and eigenvectors of Eqs. \eqref{charge_TL} and \eqref{spin_TL} to find the charge and spin velocities and their coupling while propagation.}\\

The matrix form of Eqs. \eqref{charge_TL} and \eqref{spin_TL} is given by
\begin{equation}
\begin{array}{l}
\left[ {\begin{array}{*{20}{c}}
	{{C_{eff}}}&0&0&0\\
	0&{\dfrac{C_Q}{\alpha^2}}&{p_0 g_m {L_M}}&0\\
	0&0&{{L_{eff}}}&0\\
	{ - p_0 r_m {C_{eff}}}&0&0&{\alpha^2 {L_K}}
	\end{array}} \right]\dfrac{\partial }{{\partial t}}\left\{ {\begin{array}{*{20}{c}}
	{{V_c}}\\	{{V_s}}\\	{{I_c}}\\	{{I_s}}
	\end{array}} \right\} \\+ \left[ {\begin{array}{*{20}{c}}
	0&0&0&0\\
	0&{{G_{sh}}}&{ - {p_0}{\gamma _s}}&0\\
	0&{ - {p_0}{\eta _c}}&{{R_c}}&0\\
	{ - {p_0}{\eta _s}}&0&0&{{R_s}}
	\end{array}} \right]\left\{ {\begin{array}{*{20}{c}}
	{{V_c}}\\	{{V_s}}\\	{{I_c}}\\	{{I_s}}
	\end{array}} \right\}\\ =  - \left[ {\begin{array}{*{20}{c}}
	0&0&1&0\\
	0&0&0&1\\
	1&0&0&0\\
	0&1&0&0
	\end{array}} \right]\dfrac{\partial }{{\partial x}}\left\{ {\begin{array}{*{20}{c}}
	{{V_c}}\\
	{{V_s}}\\
	{{I_c}}\\
	{{I_s}}
	\end{array}} \right\}.
\end{array}
\end{equation}

In the low-loss limit, we assume a solution of the form
\begin{equation}
 \left\{\begin{array}{*{20}{c}}
	{{V_c}}\\
	{{V_s}}\\
	{{I_c}}\\
	{{I_s}}
	\end{array} \right\}\equiv \left\{\begin{array}{*{20}{c}}
	{{\tilde{V}_c}}\\
	{{\tilde{V}_s}}\\
	{{\tilde{V}_c}}\\
	{{\tilde{V}_s}}
	\end{array} \right\} e^{j\left(kx - \omega t\right)},
\end{equation}
which results in
\begin{equation}
\label{int_step}
\begin{array}{l}
-j\omega \left[ {\begin{array}{*{20}{c}}
	{{C_{eff}}}&0&0&0\\
	0&{\dfrac{C_Q}{\alpha^2}}&{p_0 g_m {L_M}}&0\\
	0&0&{{L_{eff}}}&0\\
	{ - p_0 r_m {C_{eff}}}&0&0&{\alpha^2 {L_K}}
	\end{array}} \right] \left\{\begin{array}{*{20}{c}}
{{\tilde{V}_c}}\\
{{\tilde{V}_s}}\\
{{\tilde{V}_c}}\\
{{\tilde{V}_s}}
\end{array} \right\} \\+ \left[ {\begin{array}{*{20}{c}}
	0&0&0&0\\
	0&{{G_{sh}}}&{ - {p_0}{\gamma _s}}&0\\
	0&{ - {p_0}{\eta _c}}&{{R_c}}&0\\
	{ - {p_0}{\eta _s}}&0&0&{{R_s}}
	\end{array}} \right] \left\{\begin{array}{*{20}{c}}
{{\tilde{V}_c}}\\
{{\tilde{V}_s}}\\
{{\tilde{V}_c}}\\
{{\tilde{V}_s}}
\end{array} \right\}\\ =  -jk \left[ {\begin{array}{*{20}{c}}
	0&0&1&0\\
	0&0&0&1\\
	1&0&0&0\\
	0&1&0&0
	\end{array}} \right] \left\{\begin{array}{*{20}{c}}
{{\tilde{V}_c}}\\
{{\tilde{V}_s}}\\
{{\tilde{V}_c}}\\
{{\tilde{V}_s}}
\end{array} \right\}.
\end{array}
\end{equation}

We assume that the co-efficients of the transmission line model (Eqs. \eqref{charge_TL} and \eqref{spin_TL}) are constant with frequency $\omega$ and the propagation vector $k$. We differentiate both sides of Eq. \eqref{int_step} with respect to $k$ and the following matrix equation
\begin{equation}
\label{vel_mat}
v_g  \left\{\begin{array}{*{20}{c}}
{{\tilde{V}_c}}\\
{{\tilde{V}_s}}\\
{{\tilde{V}_c}}\\
{{\tilde{V}_s}}
\end{array} \right\} =   \left[ {\begin{array}{*{20}{c}}
	0&0&{\dfrac{1}{{{C_{eff}}}} }&0\\
	{ - \dfrac{\alpha^2 p_0 g_m L_M}{C_Q L_{eff}}}&0&0&{\dfrac{\alpha^2}{{{C_Q}}}}\\
	{\dfrac{1}{{{L_{eff}} }}}&0&0&0\\
	0&{\dfrac{1}{{\alpha^2{L_K}}}}&\dfrac{p_0 r_m}{\alpha^2L_K}&0
	\end{array}} \right]\left\{\begin{array}{*{20}{c}}
{{\tilde{V}_c}}\\
{{\tilde{V}_s}}\\
{{\tilde{V}_c}}\\
{{\tilde{V}_s}}
\end{array} \right\},
\end{equation}
where $v_g$ is the group velocity given by
\begin{equation}
v_g=\dfrac{\partial \omega }{\partial k}.
\end{equation}

Note that the eigenvalues of Eq. \eqref{vel_mat} give the velocities in Eqs. \eqref{spin_vel} and \eqref{charge_vel}. For a particular eigenvalue $v_g$, we can write the following equation from Eq. \eqref{vel_mat}
\begin{equation}
\label{eigvec}
\begin{array}{l}
\left[ {\begin{array}{*{20}{c}}
	{ - {v_{g}}}&{\dfrac{1}{{{C_{eff}}}}}\\
	{\dfrac{1}{{{L_{eff}}}}}&{ - {v_{g}}}
	\end{array}} \right]\left\{ {\begin{array}{*{20}{c}}
	{{V_c}}\\
	{{I_c}}
	\end{array}} \right\} = \left\{ {\begin{array}{*{20}{c}}
	0\\
	0
	\end{array}} \right\},\\
\left[ {\begin{array}{*{20}{c}}
	{ - {v_{g}}}&{\dfrac{1}{{{C_Q}}}}\\
	{\dfrac{1}{{{L_K}}}}&{ - {v_{g}}}
	\end{array}} \right]\left\{ {\begin{array}{*{20}{c}}
	{{V_s}}\\
	{{I_s}}
	\end{array}} \right\} = {p_0}\left\{ {\begin{array}{*{20}{c}}
	{\dfrac{{\alpha^2{g_m}{L_M}}}{{{C_Q}{L_{eff}}}}{V_c}}\\
	{ - \dfrac{{{r_m}}}{{\alpha^2{L_K}}}{I_c}}
	\end{array}} \right\}.
\end{array}
\end{equation}

For the eigenvalue $v_g=v_{g,s}={1}/{\sqrt{L_K C_Q}}$ in Eq. \eqref{eigvec} we have $V_c=I_c=0$ and assuming $I_s=1$ we get Eq. \eqref{spin_eig}. Again, for the eigenvalue $v_g=v_{g,c}={1}/{\sqrt{L_{eff} C_{eff}}}$ we get Eq. \eqref{charge_eig} assuming $I_c=1$.

\section{Derivation of semiclassical model}
\label{AppB}
\noindent
\textit{This appendix provides the derivation of Eq. \eqref{semi_1D}, starting from the Boltzmann transport equation in Eq. \eqref{BTE_1D}.}

We apply a variable transformation $\xi(x,t,\vec{p},\vec{s}) \equiv E(x,t,\vec{p},\vec{s})-\mu(x,t,\vec{p},\vec{s})$ on the left hand side of Eq. \eqref{BTE_1D}, which yields
\begin{equation}
\begin{aligned}
&\frac{\partial f}{\partial t} = \frac{\partial f}{\partial \xi} \left( \frac{\partial E}{\partial t} - \frac{\partial \mu}{\partial t} \right),\\
&v_x\frac{\partial f}{\partial x} = v_x\frac{\partial f}{\partial \xi} \left( \frac{\partial E}{\partial x} - \frac{\partial \mu}{\partial x} \right), \text{ and }\\
&F_x\frac{\partial f}{\partial p_x} = F_x\frac{\partial f}{\partial \xi} \left( \frac{\partial E}{\partial p_x} - \frac{\partial \mu}{\partial p_x}\right).
\end{aligned}
\end{equation}
We first substitute $F_x =-\dfrac{\partial E}{\partial x}$ and $v_x =\dfrac{\partial E}{\partial p_x}$. Finally, we set $\dfrac{\partial f}{\partial \xi} = \dfrac{\partial f_0}{\partial E}$ to get the left hand side of Eq. \eqref{semi_1D}.

On the right hand side of Eq. \eqref{BTE_1D}, we expand both $f$ and $f'$ into Taylor series around $f_0=1 \big / \left(1+\exp\left((E(x,\vec{p},{s})-\mu_0)/k_BT\right)\right)$ with constant electrochemical potential $\mu_0$. We set $\xi'(x,t,\vec{p'},{s'}) \equiv E(x,t,\vec{p'},{s'})-{\mu}(x,t,\vec{p'},{s'})$ and $\xi_0(x,\vec{p},{s}) \equiv E(x,\vec{p},{s})-\mu_0$ and assume that $\xi$ and $\xi'$ are close to $\xi_0$ which gives
\begin{equation}
\begin{array}{l}
f \approx {f_0} + {\left( { - \dfrac{{\partial f}}{{\partial \xi }}} \right)_{\xi  = {\xi _0}}}\left( {\xi  - {\xi _0}} \right), \text{ and}\\
f' \approx {f_0} + {\left( { - \dfrac{{\partial f}}{{\partial \xi' }}} \right)_{\xi'  = {\xi _0}}}\left( {\xi'  - {\xi _0}} \right).
\end{array}
\end{equation}
We set ${\left( {\dfrac{{\partial f}}{{\partial \xi }}} \right)_{\xi  = {\xi _0}}}={\left( {\dfrac{{\partial f}}{{\partial \xi' }}} \right)_{\xi'  = {\xi _0}}}=\dfrac{\partial f_0}{\partial E}$ Thus the right hand side of Eq. \eqref{BTE_1D} becomes
\begin{equation}
{f-{f'}} = \left(-\dfrac{\partial f_0}{\partial E}\right)\left(\xi-{\xi}'\right)=\left(\dfrac{\partial f_0}{\partial E}\right){\left(\mu-{\mu'}\right)},
\end{equation}
noting that $E(x,t,\vec{p'},s')=E(x,t,\vec{p},s)$ in the elastic scattering limit.

\section{$E$-$p$ relation}
\label{AppC}
\noindent
\textit{This appendix provides the derivation of Eq. \eqref{dE_rel2}, starting from the $E$-$p$ relation in Eq. \eqref{E_rel}.}

Differentiating Eq. \eqref{E_rel} with respect to time $t$ yields
\begin{equation}
\label{EEE_rel}
\dfrac{\partial E}{\partial t}=\left( \dfrac{\vec{p}-q\vec{A}}{m}-s \,v_0 \dfrac{\vec{p}-q\vec{A}}{|\vec{p}-q\vec{A}|} \right)\cdot\left(-q\dfrac{\partial \vec{A}}{\partial t}\right)+\dfrac{\partial U_E}{\partial t}.
\end{equation}

The velocity $\vec{v}(E)=\nabla_{\vec{p}}E$ is derived from Eq. \eqref{E_rel} as
\begin{equation}
\label{ud_vel}
\vec{v}(E)=\dfrac{\vec{p}-q\vec{A}}{m}-s\, v_0 \dfrac{\vec{p}-q\vec{A}}{|\vec{p}-q\vec{A}|}.
\end{equation}

Combining Eqs. \eqref{EEE_rel} and \eqref{ud_vel} yields Eq. \eqref{dE_rel}. Note that Eq. \eqref{ud_vel} can be written as
\begin{equation}
\label{ud_vel2}
\vec{v}(E) = \left( {\frac{{|\vec p - q\vec A|}}{m} +s\, {v_0}} \right)\frac{{\vec p - q\vec A}}{{|\vec p - q\vec A|}}.
\end{equation}
where $\dfrac{{\vec p - q\vec A}}{{|\vec p - q\vec A|}}$ is an unit vector along ${{\vec p - q\vec A}}$. From Eq. \eqref{E_rel} we get
\begin{equation}
|\vec p - q\vec A| =  -s\, m\,{v_0} \pm m\sqrt {v_0^2 + \dfrac{2\left( {{E} - U_E} \right)}{m}},
\end{equation}
noting that $s^2=1$. Thus from Eq. \eqref{ud_vel2} we get
\begin{equation}
\label{velc3}
\vec v(E) =  \pm \sqrt {v_0^2 + \frac{{2\left( {{E} - U_E} \right)}}{m}} \frac{{\vec p - q\vec A}}{{|\vec p - q\vec A|}}.
\end{equation}
Note that the magnitude of the velocity for a particular energy of interest is same for all four groups in Eq. \eqref{classification} for the Rashba Hamiltonian in Eq. \eqref{Rashba} considered here.

\section{Scattering Rates}
\label{AppE}
\noindent
\textit{This appendix provides the derivation of the scattering matrix in Eq. \eqref{scatter_mat}.}

For the group $\Re \equiv U^+$, $D^-$, $U^-$, and $D^+$, we have from Eq. \eqref{scatter}
\begin{equation}
\label{scat1}
\begin{aligned}
\tilde{S}(U^+)=& {\hat S_{U^+  \leftrightarrow {D^ - }}}\left( {\left\langle {\mu \left( {{D^ - }} \right)} \right\rangle  - \left\langle {\mu \left( U^+  \right)} \right\rangle } \right)\\
&+ {\hat S_{U^+  \leftrightarrow {U^ - }}}\left( {\left\langle {\mu \left( {{U^ - }} \right)} \right\rangle  - \left\langle {\mu \left( U^+  \right)} \right\rangle } \right)\\
&+ {\hat S_{U^+  \leftrightarrow {D^ + }}}\left( {\left\langle {\mu \left( {{D^ + }} \right)} \right\rangle  - \left\langle {\mu \left( U^+  \right)} \right\rangle } \right),\\
\end{aligned}
\end{equation}

\begin{equation}
\begin{aligned}
\label{scat2}
\tilde{S}(D^-)= &{\hat S_{D^-  \leftrightarrow {U^ + }}}\left( {\left\langle {\mu \left( {{U^ + }} \right)} \right\rangle  - \left\langle {\mu \left( D^-  \right)} \right\rangle } \right) \\
&+ {\hat S_{D^-  \leftrightarrow {U^ - }}}\left( {\left\langle {\mu \left( {{U^ - }} \right)} \right\rangle  - \left\langle {\mu \left( D^-  \right)} \right\rangle } \right)\\
&+ {\hat S_{D^-  \leftrightarrow {D^ + }}}\left( {\left\langle {\mu \left( {{D^ + }} \right)} \right\rangle  - \left\langle {\mu \left( D^-  \right)} \right\rangle } \right),
\end{aligned}
\end{equation}

\begin{equation}
\begin{aligned}
\label{scat3}
\tilde{S}(U^-)= &{\hat S_{U^-  \leftrightarrow {U^ + }}}\left( {\left\langle {\mu \left( {{U^ + }} \right)} \right\rangle  - \left\langle {\mu \left( U^-  \right)} \right\rangle } \right) \\
&+ {\hat S_{U^-  \leftrightarrow {D^ - }}}\left( {\left\langle {\mu \left( {{D^ - }} \right)} \right\rangle  - \left\langle {\mu \left( U^-  \right)} \right\rangle } \right)\\
&+ {\hat S_{U^-  \leftrightarrow {D^ + }}}\left( {\left\langle {\mu \left( {{D^ + }} \right)} \right\rangle  - \left\langle {\mu \left( U^-  \right)} \right\rangle } \right),\\
\end{aligned}
\end{equation}
and
\begin{equation}
\begin{aligned}
\label{scat4}
\tilde{S}(D^ + )= &{\hat S_{D^ +  \leftrightarrow {U^ + }}}\left( {\left\langle {\mu \left( {{U^ + }} \right)} \right\rangle  - \left\langle {\mu \left( D^ +   \right)} \right\rangle } \right) \\
&+ {\hat S_{D^ +  \leftrightarrow {D^ - }}}\left( {\left\langle {\mu \left( {{D^ - }} \right)} \right\rangle  - \left\langle {\mu \left( D^ +   \right)} \right\rangle } \right)\\
&+ {\hat S_{D^ +  \leftrightarrow {U^ - }}}\left( {\left\langle {\mu \left( {{U^ - }} \right)} \right\rangle  - \left\langle {\mu \left( D^ +   \right)} \right\rangle } \right).
\end{aligned}
\end{equation}
Eqs. \eqref{scat1}, \eqref{scat2}, \eqref{scat3}, and \eqref{scat4} together yields the scattering matrix in Eq. \eqref{scatter_mat} in the $\{\mu(U^+),\mu(D^-),\mu(U^-),\mu(D^+)\}^T$ basis.

\section{Charge and Spin Currents and Voltages}
\label{AppF}
\textit{This appendix provides the derivation of the Eq. \eqref{transform}.}

The current in any group is given by
\begin{equation}
\label{group_current}
I\left( \Re  \right) = \dfrac{q}{L} \sum\limits_{\vec p,s \in \Re } {{v_x}} \, f\left( {x,t,\vec p,s} \right).
\end{equation}
where $f$ is given by Eq. \eqref{occup_fact}. Under the linear response approximation, we can write
\begin{equation}
\label{lin_resp}
f\left( {x,t,\vec p,s} \right) \approx f_0 +\left( -\frac{\partial f_{0}}{\partial E} \right) \left( \mu\left( {x,t,\vec p,s} \right) - \mu_{0} \right)
\end{equation}
where, $f_0$ is given by Eq. \eqref{eq_occup_fact} with constant electrochemical potential $\mu_{0}$. Thus from Eq. \eqref{group_current}, we can write
\begin{equation}
I\left( \Re  \right) = \frac{q}{L}\,\left( {{f_0}\sum\limits_{\vec p,s \in \Re } {{v_x}}  + \frac{{{D_0}\left( \Re  \right)}}{2}\left( \left\langle {{v_x}\mu } \right\rangle_{\vec{p},s\in\Re}  - {\mu _0} \left\langle {{v_x}} \right\rangle\right) } \right),
\end{equation}
where $D_0(\Re)$ is given by Eq. \eqref{DOS} and the averaging is defined by Eq. \eqref{th_avg}.

We assume the following
\begin{equation*}
\left\langle {{v_x}\mu } \right\rangle_{\vec{p},s\in\Re} \approx \left\langle {{v_x} } \right\rangle \left\langle {\mu } \right\rangle_{\vec{p},s\in\Re},
\end{equation*}
which results in
\begin{equation}
\label{current_group}
\begin{aligned}
I\left( \Re  \right) ={\mathop{\rm sgn}} \left( {\left\langle {{v_x}} \right\rangle } \right) {\frac{q}{h} {n_m}\left( \Re  \right)\left( {\left\langle \mu  \right\rangle_{\vec{p},s\in\Re}  - {\mu _0}} \right)} \\+ {\mathop{\rm sgn}} \left( {| {{v_x}} | } \right)\dfrac{q}{L}{f_0}\sum\limits_{\vec p,s \in \Re } {{|v_x|}},
\end{aligned}
\end{equation}
where $n_m(\Re)$ is given by Eq. \eqref{NOM}.

\subsection{Charge Current}
The charge current in the channel is given by
\begin{equation}
{I_c} = I\left( {{U^ + }} \right) + I\left( {{D^ + }} \right) + I\left( {{U^ - }} \right) + I\left( {{D^ - }} \right),
\end{equation}
which in conjunction to Eq. \eqref{current_group} yields
\begin{equation}
\begin{aligned}
{I_c} = \dfrac{q}{h}\,\left[ {\dfrac{h}{L}{f_0}\sum\limits_{\vec p,s \in {U^ + }} {{|v_x|}} + M\left( {\mu \left( {{U^ + }} \right) - {\mu _0}} \right)} \right] \\+ \dfrac{q}{h}\,\left[ {\dfrac{h}{L}{f_0}\sum\limits_{\vec p,s \in {D^ + }} {{|v_x|}}  + N\left( {\mu \left( {{D^ + }} \right) - {\mu _0}} \right)} \right]\\
- \dfrac{q}{h}\,\left[ {\dfrac{h}{L}{f_0}\sum\limits_{\vec p,s \in {U^ - }} {{|v_x|}}  + N\left( {\mu \left( {{U^ - }} \right)  - {\mu _0}} \right)} \right] \\- \dfrac{q}{h}\,\left[ {\dfrac{h}{L}{f_0}\sum\limits_{\vec p,s \in {D^ - }} {{|v_x|}}  + M\left( {\mu \left( {{D^ - }} \right) - {\mu _0}} \right)} \right],
\end{aligned}
\end{equation}
where $n_m(U^+)=n_m(D^-)=M$ and $n_m(U^-)=n_m(D^+)=N$. Note that $(U^+,D^-)$ and $(U^-,D^+)$ are time-reversal symmetric pairs, hence

\begin{equation}
\begin{aligned}
\label{velo_cond}
&\sum\limits_{\vec p,s \in {U^ + }} {{|v_x|}}=\sum\limits_{\vec p,s \in {D^ - }} {{|v_x|}}, \text{  and  }\\
&\sum\limits_{\vec p,s \in {U^ - }} {{|v_x|}}=\sum\limits_{\vec p,s \in {D^ + }} {{|v_x|}},
\end{aligned}
\end{equation}

Thus the expression for charge current is given by
\begin{equation}
\label{ch_ic}
I_c = \dfrac{q}{h}\left( {M\tilde{\mu} \left( {U ^+ } \right) - N\tilde{\mu} \left( {U ^- } \right) + N\tilde{\mu} \left( {D ^+ } \right) - M\tilde{\mu} \left( {D ^- } \right)} \right),
\end{equation}
where we defined $\tilde{\mu}=\mu - \mu_{0}$.

\subsection{Spin Current}
The spin current in the channel is given by
\begin{equation}
\label{spin_cur}
{\tilde{I}}_s = \dfrac{1}{\alpha} \left(I\left( {{U^ + }} \right) + I\left( {{U^ - }} \right) - I\left( {{D^ + }} \right) - I\left( {{D^ - }} \right)\right),
\end{equation}
where $\alpha=2/\pi$ is an angular averaging factor to take into account the average $\hat{z}\cdot\vec{s}$ on a half Fermi circle (see Fig. \ref{2}(a)). Eq. \eqref{spin_cur} in conjunction to Eq. \eqref{current_group} yields
\begin{equation}
\begin{aligned}
{\tilde{I}}_s = \dfrac{1}{\alpha} \dfrac{q}{h}\,\left[ {\dfrac{h}{L}{f_0}\sum\limits_{\vec p,s \in {U^ + }} |{{v_x}}|  + M\left( {\mu \left( {{U^ + }} \right) - {\mu _0}} \right)} \right] \\- \dfrac{1}{\alpha} \dfrac{q}{h}\,\left[ {\dfrac{h}{L}{f_0}\sum\limits_{\vec p,s \in {U^ - }} |{{v_x}}|  + N\left( {\mu \left( {{U^ - }} \right) - {\mu _0}} \right)} \right]\\
- \dfrac{1}{\alpha} \dfrac{q}{h}\,\left[ {\dfrac{h}{L}{f_0}\sum\limits_{\vec p,s \in {D^ + }} |{{v_x}}|  + N\left( {\mu \left( {{D^ + }} \right) - {\mu _0}} \right)} \right] \\+ \dfrac{1}{\alpha} \dfrac{q}{h}\,\left[ {\frac{h}{L}{f_0}\sum\limits_{\vec p,s \in {D^ - }} |{{v_x}}|  + M\left( {\mu \left( {{D^ - }} \right) - {\mu _0}} \right)} \right].
\end{aligned}
\end{equation}

Applying the condition in Eq. \eqref{velo_cond}, we have the expression for channel spin current as
\begin{equation}
\label{ch_is_eq}
\begin{aligned}
{\tilde{I}}_{s} = \dfrac{1}{\alpha} \dfrac{q}{h}\left({M\tilde{\mu} \left( {U^+ } \right) - N\tilde{\mu} \left( {U^- } \right) - N\tilde{\mu} \left( {D^+ } \right) + M\tilde{\mu} \left( {D^- } \right)} \right)\\
+ \dfrac{2}{\alpha} \dfrac{q}{L}f_0 \left(\sum\limits_{\vec p,s \in {U^ + }} |{{v_x}}| - \sum\limits_{\vec p,s \in {U^ - }} |{{v_x}}|\right)
\end{aligned}
\end{equation}
Note first term is zero at equilibrium condition $\mu(U^+)=\mu(U^-)=\mu(D^+)=\mu(D^-)=\mu_{0}$.

However, the second term is non-zero even at equilibrium since
\begin{equation*}
\sum\limits_{\vec p,s \in {U^ + }} |{{v_x}}| \neq \sum\limits_{\vec p,s \in {U^ - }} |{{v_x}}|
\end{equation*}
and represent the equilibrium spin current in the channel. We subtract the equilibrium part from our definition of spin current given by
\begin{equation}
\label{ch_is}
\begin{aligned}
{{I}}_{s} = \dfrac{1}{\alpha} \dfrac{q}{h}\left({M\tilde{\mu} \left( {U^+ } \right) - N\tilde{\mu} \left( {U^- } \right) - N\tilde{\mu} \left( {D^+ } \right) + M\tilde{\mu} \left( {D^- } \right)} \right).
\end{aligned}
\end{equation}

\subsection{Spin Voltage}
The spin voltage in the channel is given by
\begin{equation}
\label{vs}
q\,V_s = \alpha \frac{{M\mu \,(U + ) + N\mu \,(U - ) - N\mu \,(D + ) - M\mu \,(D - )}}{{2\,(M + N)}}.
\end{equation}

Subtracting each electrochemical potential by $\mu_{0}$ which gives
\begin{equation}
\label{ch_vs}
q\,V_s = \alpha \frac{{M\tilde{\mu} (U + ) + N\tilde{\mu} (U - ) - N\tilde{\mu} (D + ) - M\tilde{\mu} (D - )}}{{2\,(M + N)}}.
\end{equation}

\subsection{Charge Voltage}
The charge voltage in the channel is given by
\begin{equation}
q\,\tilde{V}^c = \;\dfrac{{M\mu (U + ) + N\mu (U - ) + N\mu (D + ) + M\mu (D - )}}{{2\,(M + N)}}.
\end{equation}

Subtracting each electrochemical potential by $\mu_{0}$ according to
\begin{equation}
\label{ch_vc}
q\,V^c = \;\dfrac{{M\tilde{\mu} (U + ) + N\tilde{\mu} (U - ) + N\tilde{\mu} (D + ) + M\tilde{\mu} (D - )}}{{2\,(M + N)}}.
\end{equation}
where $V_c = \tilde{V}_c - \dfrac{\mu_{0}}{q}$.

Eqs. \eqref{ch_ic}, \eqref{ch_is}, \eqref{ch_vs}, and \eqref{ch_vc} together can be written in a matrix form given by Eq. \eqref{transform}.

\bibliography{Ref}

\end{document}